\documentclass[fleqn,usenatbib]{mnras}

\usepackage{newtxtext,newtxmath}

\usepackage[T1]{fontenc}

\usepackage{graphicx}	
\usepackage{amsmath}	

\usepackage{xcolor}

\title[Cosmology with multiple halo sparsities]{Forecasting cosmological parameter constraints using multiple sparsity measurements as tracers of the mass profiles of dark matter haloes}

\author[P.S. Corasaniti et al.]{
P.S. Corasaniti$^{1,2}$\thanks{E-mail: Pier-Stefano.Corasaniti@obspm.fr}, A.M.C. Le Brun,$^{1,3}$\thanks{E-mail: Amandine.Le-Brun@obspm.fr},
T.R.G. Richardson$^{1}$,
Y. Rasera$^{1}$,
S. Ettori$^{4,5}$, M. Arnaud$^{3}$, \newauthor
G. W. Pratt$^{3}$
\\
$^{1}$ Laboratoire Univers et Th\'eorie, Observatoire de Paris, Universit\'e PSL, Universit\'e Paris Cit\'e, CNRS, F-92190 Meudon, France\\
$^{2}$ Sorbonne Universit\'e, CNRS, UMR 7095, Institut d'Astrophysique de Paris, 98 bis bd Arago, 75014 Paris, France\\
$^{3}$ AIM, CEA, CNRS, Universit\'e Paris-Saclay, Universit\'e Paris Cit\'e, Sorbonne Paris Cit\'e, F-91191 Gif-sur-Yvette, France\\
$^{4}$ INAF, Osservatorio di Astrofisica e Scienza dello Spazio di Bologna, via Piero Gobetti 93/3, I-40129 Bologna, Italy \\
$^{5}$ INFN, Sezione di Bologna, viale Berti Pichat 6/2, 40127 Bologna, Italy \\
}

\pubyear{2022}

\begin{document}
\label{firstpage}
\pagerange{\pageref{firstpage}--\pageref{lastpage}}
\maketitle

\begin{abstract}
The dark matter halo sparsity, i.e. the ratio between spherical halo masses enclosing two different overdensities, provides a non-parametric proxy of the halo mass distribution which has been shown to be a sensitive probe of the cosmological imprint encoded in the mass profile of haloes hosting galaxy clusters. Mass estimations at several overdensities would allow for multiple sparsity measurements, that can potentially retrieve the entirety of the cosmological information imprinted on the halo profile. Here, we investigate the impact of multiple sparsity measurements on the cosmological model parameter inference. For this purpose, we analyse N-body halo catalogues from the Raygal and M2Csims simulations and evaluate the correlations among six different sparsities from Spherical Overdensity halo masses at $\Delta=200,500,1000$ and $2500$ (in units of the critical density). Remarkably, sparsities associated to distinct halo mass shells are not highly correlated. This is not the case for sparsities obtained using halo masses estimated from the Navarro-Frenk-White (NFW) best-fit profile, that artificially correlates different sparsities to order one. This implies that there is additional information in the mass profile beyond the NFW parametrisation and that it can be exploited with multiple sparsities. In particular, from a likelihood analysis of synthetic average sparsity data, we show that cosmological parameter constraints significantly improve when increasing the number of sparsity combinations, though the constraints saturate beyond four sparsity estimates. We forecast constraints for the CHEX-MATE cluster sample and find that systematic mass bias errors mildly impact the parameter inference, though more studies are needed in this direction. 
\end{abstract}




\section{Introduction}

At the time of writing, a heavy focus within the field of precision cosmology is set on constraining the parameters of the $\Lambda$CDM cosmological model, named according to its two main constituents: a cosmological constant $\Lambda$ and Cold Dark Matter (CDM), while also exploring possible extensions to this model by investigating alternative scenarios of dark matter \citep[see e.g.][]{Boyarsky2019,Niemeyer2020,Green2020} and dark energy \citep[see e.g.][]{CopelandTsujikawa2006,Brax2018}.

To this effect, many probes have been devised and applied to a range of observations such as Cosmic Microwave Background (CMB) experiments \citep[e.g ][]{Fixsen1996,Komatsu2011,Planck2020}, Big Bang Nucleo-synthesis estimates \cite[e.g.][]{Aver2015,Cooke2018}, Large Scale Structure (LSS) observations through measurements of the clustering of matter measured with various probes, such as galaxies \citep[see e.g.][]{2001MNRAS.327.1297P,2004ApJ...606..702T,2005MNRAS.362..505C,2017MNRAS.466.2242B} or Lyman-$\alpha$ absorbing gas \citep{2016MNRAS.457.3541C}, gravitational lensing \citep[e.g.][]{Wong2020, Birrer2020,Gatti2021}, measurements of the Baryonic Acoustic Oscillation (BAO) scale \citep[e.g.][]{destAgathe2019,deMattia2021, desY3bao2021} and tests of the Hubble diagram from SN Ia standard-candles  \cite[e.g.][]{Anand2021,Riess2021} to cite but a few examples.

In this regard, galaxy clusters have proven to be a useful asset. These structures, the most massive gravitationally bound in the Universe, exhibit multiple properties that can be used to test the cosmological paradigm. As an example, cosmological constraints have been inferred from measurements of their abundance \citep[e.g.][]{PlanckSZ2014,PlanckSZ2016,Pacaud2018,Bocquet2019, Lesci2020,To2021}, their spatial clustering \citep{2014A&A...571A..21P,Maruli2021} or the fraction of gas contained inside their potential wells \citep[e.g.][]{Ettori2003,Allen2008,Ettori2009,Mantz2014, Mantz2022}. 

The possibility of extracting cosmological information from estimates of the mass profiles of galaxy clusters is another probes that however remains relatively unexplored \citep[see e.g.][]{2010A&A...524A..68E}. Historically, this approach has been built upon the remarkable result that density profiles of dark matter haloes from N-body simulation are described to a good approximation by a two-parameter universal function, the Navarro-Frenk-White profile \citep[NFW;][]{NFW1997}. However, the difficulty of obtaining accurate estimates of the concentration-mass relations from galaxy cluster observations has so far been the main limitation to the use of cluster mass profiles as cosmological proxy \citep[see e.g.][]{2010MNRAS.406..434M,2011MNRAS.416.2539K,2015MNRAS.449.2024S}.

In recent years, a novel approach in this direction has been developed around the concept of halo \textit{sparsity}, that is the ratio of halo masses measured at radii enclosing different overdensities, as a non-parametric proxy for the internal halo mass distribution. In the seminal work of \citet{Balmes2014}, it has been shown that halo sparsity depends on the characteristics of the underling cosmological model. Further investigation by \citet{Corasaniti2020} has found that the average sparsity is also sensitive to modified gravity scenarios and can therefore be used to constrain the latter. Recently, cosmological constraints using measurements of the average halo sparsity of galaxy cluster samples have yielded results competitive with other widely used probes \citep{Corasaniti2018, Corasaniti2021}.

The average halo sparsity has been shown to possess a number of interesting features \citep[see e.g.][]{Corasaniti2018,Corasaniti2019,Corasaniti2021}. On the one hand, it provides a simple link between measurements of the mass profile of an ensemble of galaxy clusters and cosmological model predictions derived from an integral relation involving the halo mass function at the overdensities of interest. On the other hand, being a mass ratio, the average sparsity is less impacted by the systematic errors known to affect the measurements of galaxy cluster masses \citep[see e.g.][]{2007ApJ...655...98N,2010A&A...514A..93M,2012NJPh...14e5018R,2014MNRAS.442.2641V,2016ApJ...827..112B,2015MNRAS.450.3633S}. Furthermore, the properties that characterise the halo sparsity are independent of the specific form of the halo density profile. As such, the use of multiple sparsity measurements from non-parametric mass estimates opens the way to retrieving cosmological information encoded over the entire halo mass profile rather than from a single determination at two particular overdensities. However, because of the gravitational assembly processes shaping the mass distribution of haloes, we can expect these different sparsities to be correlated. 

Here, we set to evaluate the minimum number of multiple average sparsity estimates that sample the halo mass profile at different overdensities, while providing maximal constraints on a set of cosmological parameters. For this purpose, we perform a thorough analysis of average sparsities and their correlations using halo catalogues from large volume high-resolution N-body simulations. Building upon this numerical study, we perform a Markov Chain Monte Carlo likelihood analysis on synthetic datasets to investigate the level of cosmological parameter constraints that can be inferred from different combinations of average sparsity measurements under different fiducial cosmologies and cluster mass measurement error assumptions. 

The paper is organised as follows: in Section~\ref{spars_def}, we introduce the basic concepts, describe the N-body simulations and present the results of the analysis of numerical halo catalogues; in Section~\ref{mcmc_analysis}, we describe the cosmological parameter inference from multiple  average sparsity measurements for two distinct synthetic datasets, while in Section~\ref{forecast}, we present a parameter forecast analysis for a realistic galaxy cluster sample. Finally, in Section~\ref{conclusions}, we discuss the conclusions.

\section{Cosmology with Halo Sparsity}\label{spars_def}
\subsection{Definition \& Properties}
\begin{figure}
    \centering
    \includegraphics[width = 0.9\linewidth]{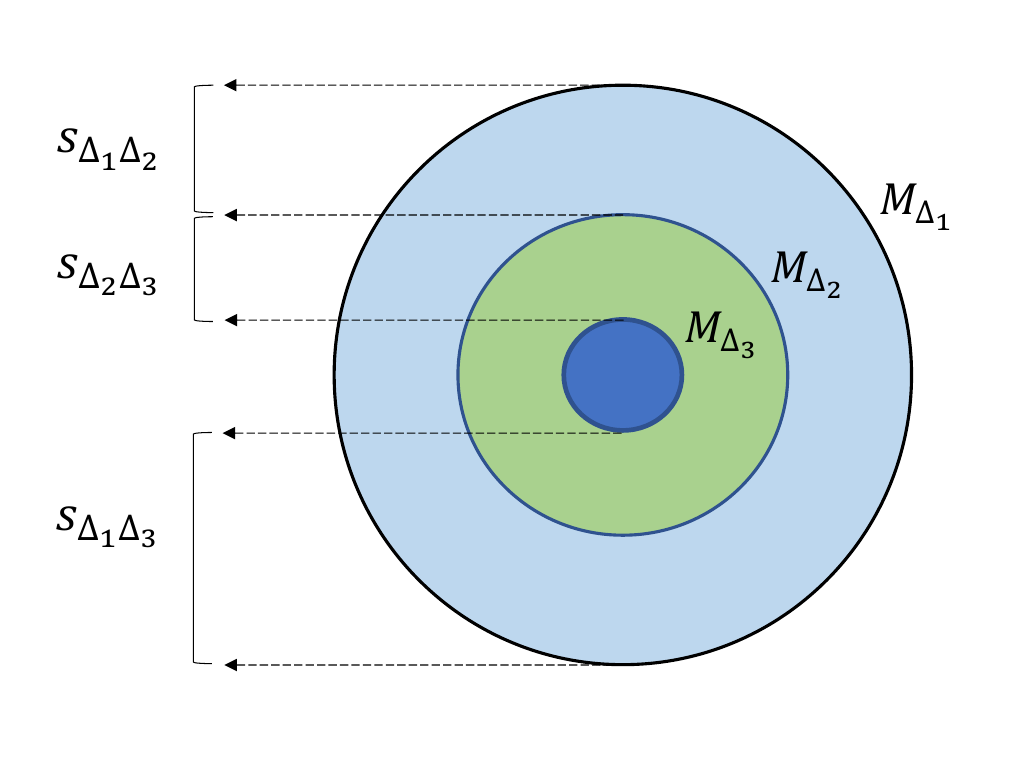}
    \caption{2D illustration of the spherical mass profile of a halo as probed by multiple sparsity estimates at overdensities $\Delta_1<\Delta_2<\Delta_3$. Each sparsity $s_{\Delta_i,\Delta_j}$ measures the fractional mass within the spherical shell comprised between $r_{\Delta_i}$ and $r_{\Delta_j}$ relative to the mass enclosed in the inner radius $r_{\Delta_j}$.}
    \label{fig:halo_spars}
\end{figure}

Halo sparsity is defined as \citep{Balmes2014}:
\begin{equation}
    s_{\Delta_1,\Delta_2}=\frac{M_{\Delta_1}}{M_{\Delta_2}},
\end{equation}
where $M_{\Delta_1}$ and $M_{\Delta_2}$ are halo masses at radii $r_{\Delta_1}$ and $r_{\Delta_2}$ which enclose the overdensity $\Delta_1$ and $\Delta_2$ respectively, with $\Delta_1<\Delta_2$ (with the overdensities in units of the critical $\rho_c$ or background $\rho_b$ density). This  ratio can also be interpreted as the ratio of the mass $\Delta{M}_{12}$ within the radial shell $\Delta{r}=r_{\Delta_1}-r_{\Delta_2}$ and of the mass within the inner radius $r_{\Delta_2}$, i.e. $s_{\Delta_1,\Delta_2}=\Delta{M}_{12}/M_{\Delta_2}+1$. Hence, the values of sparsities at multiple overdensity pairs probe the fractional mass profile of the halo. As an example, in Fig.~\ref{fig:halo_spars}, we show a graphic illustration of the case of halo masses at overdensities $\Delta_1$, $\Delta_2$ and $\Delta_3$, which allow to estimate three sparsity combinations $s_{\Delta_1,\Delta_2}$, $s_{\Delta_1,\Delta_3}$ and $s_{\Delta_2,\Delta_3}$.

Quite importantly, at any given redshift, the halo sparsity is largely independent of the outer halo mass $M_{\Delta_1}$ \citep{Balmes2014,Corasaniti2018,Corasaniti2019}; consequently, for a given pair of overdensities, the ensemble average value can be obtained by integrating the equality
\begin{equation}
    \frac{dn}{dM_{\Delta_2}}=\frac{dn}{dM_{\Delta_1}}s_{\Delta_1,\Delta_2}\frac{d\ln{M_{\Delta_1}}}{d\ln{M_{\Delta_2}}},
\end{equation}
where $dn/dM_{\Delta_2}$ is the mass function at $M_{\Delta_2}$ of the ensemble of haloes with mass function $dn/dM_{\Delta_1}$ at $M_{\Delta_1}$ (i.e. mass functions of matched haloes), to obtain the average sparsity relation:
\begin{equation}\label{spars_mf}
    \int_{M_{\Delta_2}^{\rm min}}^{M_{\Delta_2}^{\rm max}}\frac{dn}{dM_{\Delta_2}}d\ln{M_{\Delta_2}}=\langle s_{\Delta_1,\Delta_2}\rangle\int_{\langle s_{\Delta_1,\Delta_2}\rangle M_{\Delta_2}^{\rm min}}^{\langle s_{\Delta_1,\Delta_2}\rangle M_{\Delta_2}^{\rm max}}\frac{dn}{dM_{\Delta_1}}d\ln{M_{\Delta_1}}.
\end{equation}
Given the functional form of $dn/dM_{\Delta_1}$ and $dn/dM_{\Delta_2}$ the above equation can be solved numerically to obtain the value of the average sparsity. \citet{Corasaniti2018} has shown that Eq.~(\ref{spars_mf}) provides predictions of the average sparsity that are accurate to a few percent level for $s_{200,500}$ and $s_{500,1000}$, thus providing the foundations to perform cosmological parameter inference using average sparsity measurements. 

Notice that Eq.~(\ref{spars_mf}) is largely insensitive to the choice of the integration limits. Indeed, since at the high-mass end the mass function drops exponentially, the upper limit can be set to any arbitrary large number; while at the low-mass end, given that the halo sparsity is nearly constant as a function of halo mass, the integral can be set without loss of generality to the minimum halo mass of the halo catalogues used for the calibration of the mass functions.

Hereafter, we will test the validity of Eq.~(\ref{spars_mf}) over a wider range of overdensities than those originally investigated in \citet{Corasaniti2018}, which is a necessary step to infer cosmological parameter constraints from multiple average sparsity determinations.

\subsection{N-body Simulations}\label{nbody}
We use halo catalogues from two distinct sets of N-body simulations, characterised by different cosmological model parameters, but approximately similar mass resolution and generated with the same simulation code. This enables us to extend our investigation of the average sparsity correlations to the dependence upon the underlying cosmological model (around the $\Lambda$CDM model best-fit to the CMB data). 

\subsubsection{RayGalGroupSims}
The RayGalGroupSims $\Lambda$CDM simulation, or simply Raygal, consists of a ($2.6$ Gpc $h^{-1}$)$^3$ volume and sampled with $4096^3$ particles (corresponding to a particle mass resolution $m_p=1.88\times 10^{10}\,M_{\odot}\,h^{-1}$) realised with the Adaptive Mesh Refinement (AMR) N-body code RAMSES \citep{Teyssier2002}. The cosmological model parameters have been set consistently to the WMAP-7 year data analysis of a flat $\Lambda$CDM model \citep{Komatsu2011}: $\Omega_m = 0.2573$, $\Omega_b = 0.04356$, $h = 0.72$, $n_s = 0.963$ and $\sigma_8 = 0.801$. We refer interested readers to \citet{2019MNRAS.483.2671B,Rasera2021} for a detailed description of the RayGalGroupSims suite. Full redshift snapshots have been stored at $z=0.00,0.50,0.66,1.00,1.14,1.50$ and $2.00$. 

\subsubsection{M2Csims $\Lambda$CDM Simulation}
The M2Csims $\Lambda$CDM simulation suite consists of three ($1$ Gpc $h^{-1}$)$^3$ volume boxes with $2048^3$ particles (corresponding to a particle mass resolution $m_p=1.02\times 10^{10}\,h^{-1}\,M_{\odot}$) run with the Adaptive Mesh Refinement (AMR) N-body code RAMSES \citep{Teyssier2002}. The cosmological model parameters are set to the \textit{Planck}-2015 $\Lambda$CDM cosmology \citep{2016A&A...594A..13P} with $\Omega_m = 0.3156$, $\Omega_b = 0.0492$, $h = 0.6727$, $n_s = 0.9645$ and $\sigma_8 = 0.831$.  We refer interested readers to \citet{LeBrun2018} for a more detailed description of the suite (a complete description will appear in Le Brun et al. in preparation). Note that the suite also contains high-resolution zooms for more than 450 massive galaxy clusters which will not be used here as the most important requirement is the number of galaxy clusters over the resolution of their profiles. Full snapshots have been stored at $z=0.00,0.125,0.25,0.30,0.50,0.60,0.75,0.80,1.00,1.25$ and $1.50$. In the analyses presented here, we use catalogues from only two of the M2Csims simulation suite corresponding to a total comoving volume of $2$ (Gpc $h^{-1}$)$^3$, slightly smaller than that of the Raygal simulation.

\subsection{N-body Halo Catalogues}
Halo catalogues for both simulations have been generated with the Spherical Overdensity (SO) algorithm \citep{1994MNRAS.271..676L} implemented in the parallel code \texttt{pSOD}\footnote{The parallelization scheme has been adopted from the code \texttt{pFOF} \citep{2014A&A...564A..13R}.}. The algorithm first evaluates the particle density in each cell, then starts from the cell with maximum density. In each candidate cell, the centre position is chosen to be that of the particle with the greatest number of neighbouring particles within a sphere of a given radius. Afterwards, the SO algorithm computes the particle density in spheres of increasing radii around that central particle until it reaches the overdensity threshold $\Delta$. Hereafter, we will always refer to overdensities given in units of the critical density. We focus on haloes detected with an overdensity threshold $\Delta=200$. For each halo in the catalogues, we estimate masses at overdensities $\Delta=200,500,750, 1000,1500,2000$ and $2500$ respectively. In order to be exempt of numerical resolution artefacts, we further select haloes with $M_{2500c}>10^{13}$ M$_{\odot}\,h^{-1}$. This also guarantees us that we consider haloes with masses $M_{200c}>10^{13}$ M$_{\odot}\,h^{-1}$, thus corresponding to haloes hosting galaxy groups and clusters.

Since we are interested in the application of multiple average sparsity measurements to galaxy cluster observations, we limit our analysis to halo catalogues in the redshift range $0\le z\le 1.5$. This is because the detection of clusters at higher redshifts as well as the estimation of the cluster masses at the level of accuracy required seems currently unrealistic. Also, for consistency with the conventions of the galaxy cluster community, we focus on halo masses at overdensities $\Delta=200,500,1000$ and $2500$, respectively.

\subsection{Halo Mass Function Calibration}\label{hmf}
We compute the halo mass function for each of the mass overdensity definitions in the halo catalogues as:
\begin{equation}
    \frac{dn}{d\ln{M_{\Delta}}}= \frac{N(M_{\Delta})}{\Delta\ln{M_{\Delta}}}\frac{1}{L^3},
\end{equation}
where $N(M_{\Delta})$ is the number of haloes in a logarithmic mass bin of size $\Delta\ln{M_{\Delta}}=0.3$ centred at $M_{\Delta}$ and $L$ is the size of the simulation box. We use the numerical estimates of the halo mass functions to calibrate, at each redshift snapshot, the coefficients of the Sheth-Tormen \citep[ST;][]{1999MNRAS.308..119S} formula $f_{\rm ST}$ as given by:
\begin{equation}
    \frac{dn}{dM_{\Delta}}=\frac{\rho_m}{M_{\Delta}}\left(-\frac{1}{\sigma}\frac{d\sigma}{dM_{\Delta}}\right)f_{\rm ST}(\sigma),
\end{equation}
where $\rho_m$ is the cosmic matter density, $\sigma(M_\Delta)$ is the root-mean-square fluctuation of the linear density field smoothed on a scale enclosing the mass $M_{\Delta}$ and 
\begin{equation}\label{st_multiplicity}
    f_{\rm ST}(\sigma) = A_{\Delta}\frac{\delta_c}{\sigma}\sqrt{\frac{2 a_{\Delta}}{\pi}}\left[1+\left(\frac{a_{\Delta}\delta_c^2}{\sigma^2}\right)^{-p_{\Delta}}\right]e^{-\frac{a_{\Delta}\delta_c^2}{2\sigma^2}},
\end{equation}
where $A_{\Delta}$, $a_{\Delta}$ and $p_{\Delta}$ are calibration parameters and $\delta_c$ is the linearly extrapolated spherical collapse threshold, which we compute using the formula by \citet{KitayamaSuto1996}. 

It is worth noticing that the functional form of the Sheth-Tormen parametrization is the base of all the numerically calibrated formula that aim to predict the halo mass function for any given set of cosmological parameters \citep[see e.g.][]{2008ApJ...688..709T,2016MNRAS.456.2361B,2016MNRAS.456.2486D,2021MNRAS.500.2316C}. This is because such a form of multiplicity function manifests a high level of universality. Here, we have explicitly kept the dependence on linear spherical collapse threshold $\delta_c$, which as shown in \citet{2011MNRAS.410.1911C}, it allows to better account for the cosmology dependence of the multiplicity function. On the other hand, a key difference among the calibrated multiplicity functions discussed in the literature concerns the redshift parametrisation of the ST parameters. Here, we follow the approach of \citet{2016MNRAS.456.2486D} and parametrise the redshift dependence of the ST coefficients in terms of an expansion in logarithmic powers of the overdensity $\Delta$ relative to the virial overdensity at the redshift of interest. The intent is to capture the redshift evolution of the halo mass function at different overdensities $\Delta$ for the sample of halos detected at $\Delta=200$. In particular, we assume a quadratic expansion and we refer the reader to Appendix~\ref{app_a} for a detailed description of the fitting procedure of the halo mass functions. Nonetheless, we would like to stress that our calibration substantially differs from that of \citet{2016MNRAS.456.2486D}, who have provided fitting formula calibrated on halo samples detected with different overdensities. As such, their mass function cannot be used to predict the halo sparsity unless corrections are taken into account for the systematic due to effect of unmatched halos as described in \citep{Corasaniti2021}. 

Alternatively, the halo mass function for a given cosmological setup can be predicted from emulators. These are built using halo catalogues from suites of N-body simulations with different cosmological parameters \citep[see e.g.][]{2019ApJ...872...53M,2019ApJ...884...29N,2020ApJ...901....5B}. In a similar manner, it should be possible to build emulators of the average halo sparsity, a possibility which we will investigate in a future study.

\begin{figure*}
    \includegraphics[width = 0.5\linewidth]{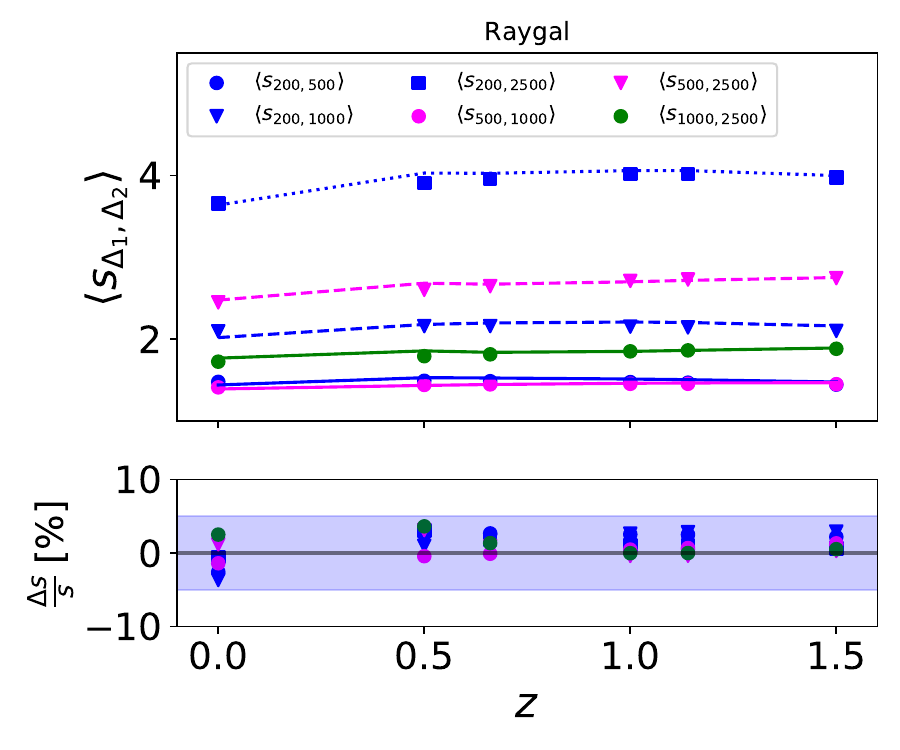}\includegraphics[width = 0.5\linewidth]{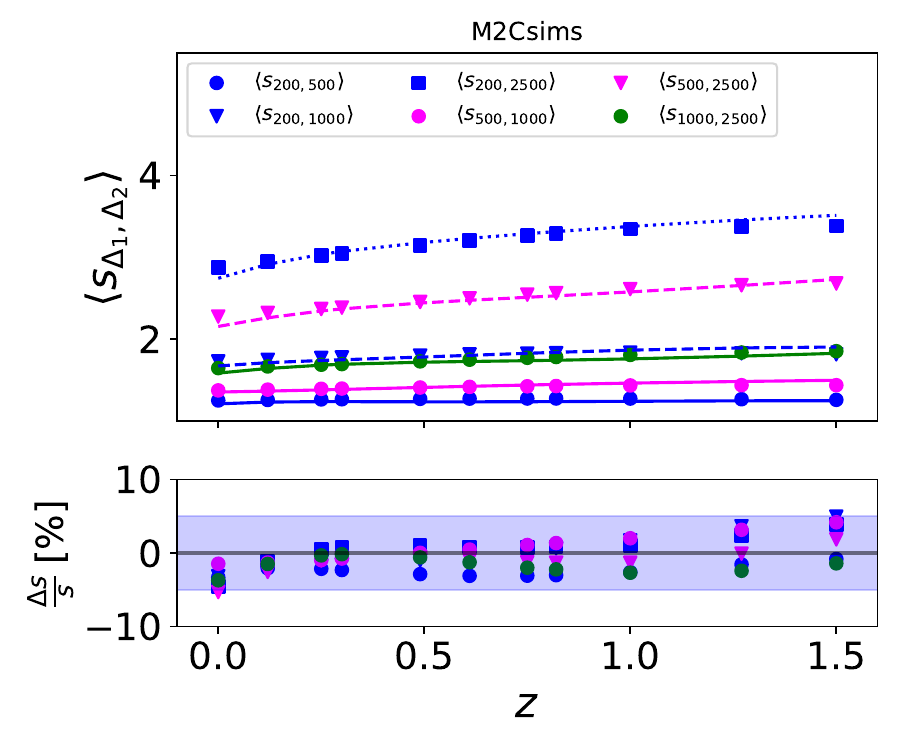}
    \caption{Average halo sparsity estimates for different overdensity configurations as a function of redshift for the Raygal (left panel) and M2Csims (right panel) halo catalogues. The data points in the plots correspond to the N-body estimates, while the various lines show the analytical predictions from the average sparsity relation of Eq.~(\ref{spars_mf}) that has been solved using the ST-parametrised mass functions calibrated on the simulations. The relative differences between the N-body and analytical results are shown in the bottom panels, where the shaded areas delimit the regions with less than 5 per cent relative differences. Differences between the average sparsity values from the Raygal and M2Csims halo catalogues for the same overdensity pairs pertain to the differences of the simulated cosmological models.}
    \label{fig:av_spars}
\end{figure*}

\subsection{Average Sparsities}\label{average_spars}

Given a set of mass estimates $M_{\Delta_i}$ measured at $n$ overdensities $\Delta_i$ (in units of the critical density), we can compute up to $N_s =\binom{n}{2}=\frac{n!}{2(n-2)!}$ distinct sparsities. As such, if we consider a number $m$ of them with $m<N_s$, then the number of possible $m$ sparsity combinations is given by $N_m=\binom{N_s}{m}=\frac{N_s!}{m!(N_s - m)!}$.

As already mentioned, here we restrict ourselves to $n = 4$ mass measurements at $\Delta=200,500,1000$ and $2500$ and for each halo in the catalogues, we focus on the following set of $N_s = 6$ halo sparsities: $s_{200,500}$, $s_{200,1000}$, $s_{200,2500}$, $s_{500,1000}$, $s_{500,2500}$ and $s_{1000,2500}$. Notice that in such a case, there is a total of $N_{\rm tot}=\sum_{m=1}^{N_s}N_m=63$ possible permutations for any number $m$ of sparsities used in the analysis. The factorial dependence of the number of combinations prohibits the full exploration of this parameter space; as such, in later sections, we will clearly quote which combinations are used. From this set, at each redshift snapshot, we evaluate the halo ensemble average sparsities by computing the arithmetic mean of the individual halo sparsities:
\begin{equation}
    \langle s_{\Delta_1,\Delta_2}\rangle\equiv \left\langle\frac{M_{\Delta_1}}{M_{\Delta_2}}\right\rangle=\frac{1}{N_{h}}\sum_i^{N_{h}}s^i_{\Delta_1,\Delta_2},
\end{equation}
where $N_{h}$ is the total number of haloes in a catalogs at a given redshift.

In Fig.~\ref{fig:av_spars}, we plot the average halo sparsities for the Raygal (left panel) and M2Csims (right panel) halo catalogues respectively. In the same plots, we also show the values predicted by the solutions of Eq.~(\ref{spars_mf}) and the relative differences with respect to the N-body estimates (bottom panels). We can see that differences are well within $5\%$ level and in some cases even at sub-percent level. In Appendix~\ref{app_a}, we also show the relative differences between the prediction from the Raygal calibrated mass functions for the M2Csims cosmology and the M2Csims average halo sparsity and viceversa, which we find to be $\lesssim 5\%$, consistent with those shown in Fig.~\ref{fig:av_spars}. 

Notice that there is a systematic difference between the values of the average sparsities obtained from the Raygal simulation and those from the M2Csims case. This pertains to the cosmological dependence of the sparsity originally pointed out in \citet{Balmes2014}. 

The average halo sparsity is mainly sensitive to a degenerate combination of $\Omega_m$ and $\sigma_8$ as given by $S_8=\sigma_8\sqrt{\Omega_m/0.3}$ \citep{Corasaniti2018,Corasaniti2021}. In particular, the lower the level of clustering of a given cosmological model, i.e. the smaller the value of $S_8$, the higher the value of the average halo sparsity. This is because cosmic structures will form later in a cosmological model with lower $S_8$ than in a model with a larger value. Consequently, such structures will be less concentrated or equivalently more sparse than in a model with a larger value of $S_8$.

Comparing the trends in Fig.~\ref{fig:av_spars}, we can see that the Raygal average sparsities at any given redshift are systematically larger than the M2Csims values, which is consistent with the fact that the Raygal $\Lambda$CDM model has $S_8=0.742$, while in the case of the M2Csims we have $S_8=0.852$. It is also worth noticing that, among the different sparsity estimates, the one with the largest value and the largest variation with the underlying cosmology is associated to $\langle s_{200,2500}\rangle$. In particular, the maximum relative variation of $\langle s_{200,2500}\rangle$ with respect to the M2Csims case amounts to $\sim 20\%$. This is consistent with the results of 
\citet{Balmes2014}, who have found that the cosmological dependence of the sparsity increases as the difference between $\Delta_1$ and $\Delta_2$ increases. On the other hand, we can also notice that sparsities for different overdensity pairs do not have the same sensitivity to the underlying cosmology. As an example, we find that $\langle s_{500,1000}\rangle$ and $\langle s_{1000,2500}\rangle$ vary by only a few percent.

\begin{figure*}
    \centering
    \includegraphics[width = 1.0\linewidth]{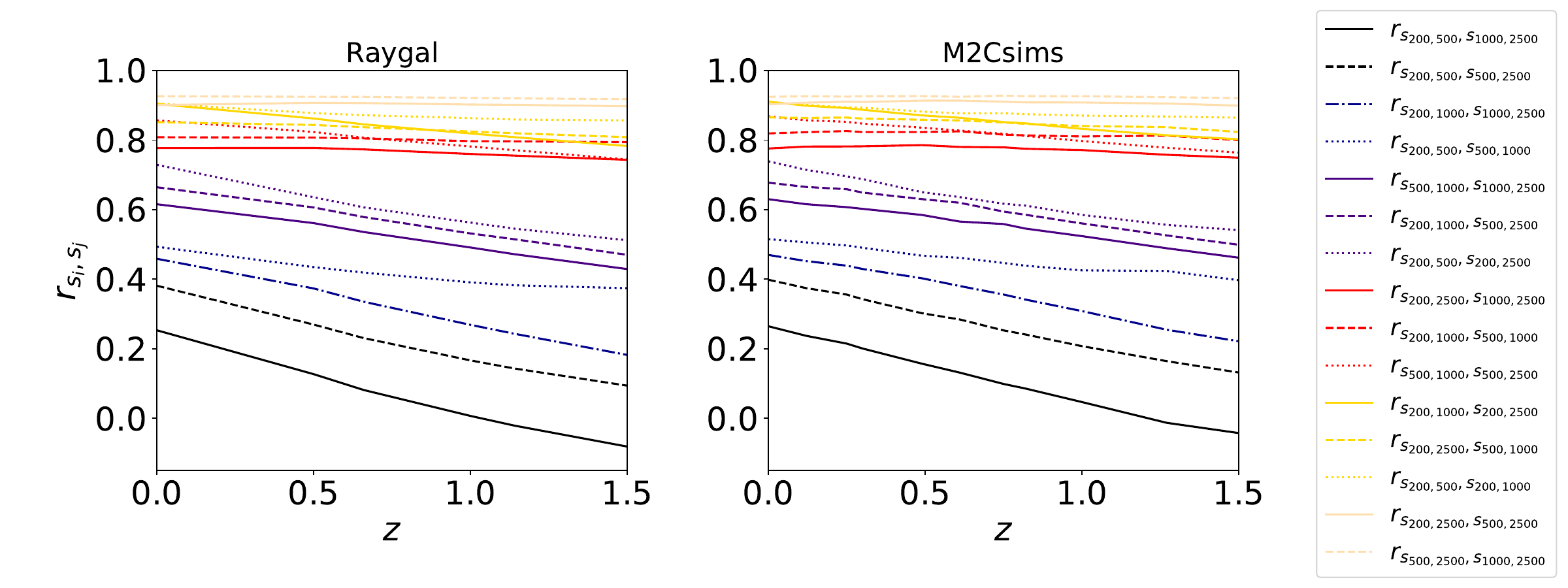}
    \caption{Sparsity correlation coefficients from the Raygal (left panel) and M2Csims (right panel) halo catalogues, respectively. As we can see over the redshift range $0<z<1.5$, sparsity combinations probing the halo mass distribution in close (overlapping) mass shells are highly correlated ($r\gtrsim 0.5$). This is not the case of sparsities associated to mass shells that are at larger separations ($r\lesssim 0.5$).}
    \label{fig:corr}
\end{figure*}

\subsection{Average Sparsities Correlations}\label{av_corr}

The cosmological information encoded in the estimated average sparsities is not independent. In fact, the gravitational processes that shape the mass assembly of the haloes correlate the properties of the mass distribution within different radial shells. For this reason, we use the data from the N-body halo catalogues to compute the correlation coefficients of the different sparsity estimates, which is given by:
\begin{equation}\label{corr_coeff}
    r_{s_i,s_j}=\frac{\sum_{k=1}^{N_{\rm h}}\left(s_i^k-\langle s_i\rangle \right)\left(s_j^k-\langle s_j\rangle\right)}{\sqrt{\sum_{k=1}^{N_{\rm h}}\left(s_i^k-\langle s_i\rangle\right)^2\sum_{k=1}^{N_{\rm h}}\left(s_j^k-\langle s_j\rangle\right)^2}},
\end{equation}
where the index $i,j=\left\{(200,500),(200,1000),...,(1000,2500)\right\}$ with $i\ne j$. They are shown in Fig.~\ref{fig:corr} as a function of redshift for the Raygal and M2Csims halo catalogues, respectively. In order to facilitate the visualization of the low correlated pairs of sparsity configurations against the highly correlated one we have adopted the \textit{magma} colormap for the colours of the various lines.

First of all, from Fig.~\ref{fig:corr}, we may notice that all correlations increase from high to low redshifts both for the Raygal haloes and M2Csims ones. This is a direct consequence of the mass assembly process of haloes, which grow from inside out \citep{2011MNRAS.413.1373W,2011AdAst2011E...6T}. As the haloes assemble their mass over cosmic time, the mass distributions within different mass shells become increasingly correlated. Secondly, we can see that the correlations are smaller for sparsities that sample the mass profile within mass shells that are at larger separations. As an example, $s_{200,500}$ and $s_{1000,2500}$ have a maximal $\sim 25$ per cent correlation at $z=0$, which is not the case for sparsities probing the mass distribution in close mass shells (or even overlapping ones) with correlation greater than $50$ per cent. We find the redshift evolution of the correlation coefficients to be well approximated by a linear regression, which we provide in Appendix \ref{app_b} for practical applications.

\begin{figure}
    \centering
    \includegraphics[width = 0.95\linewidth]{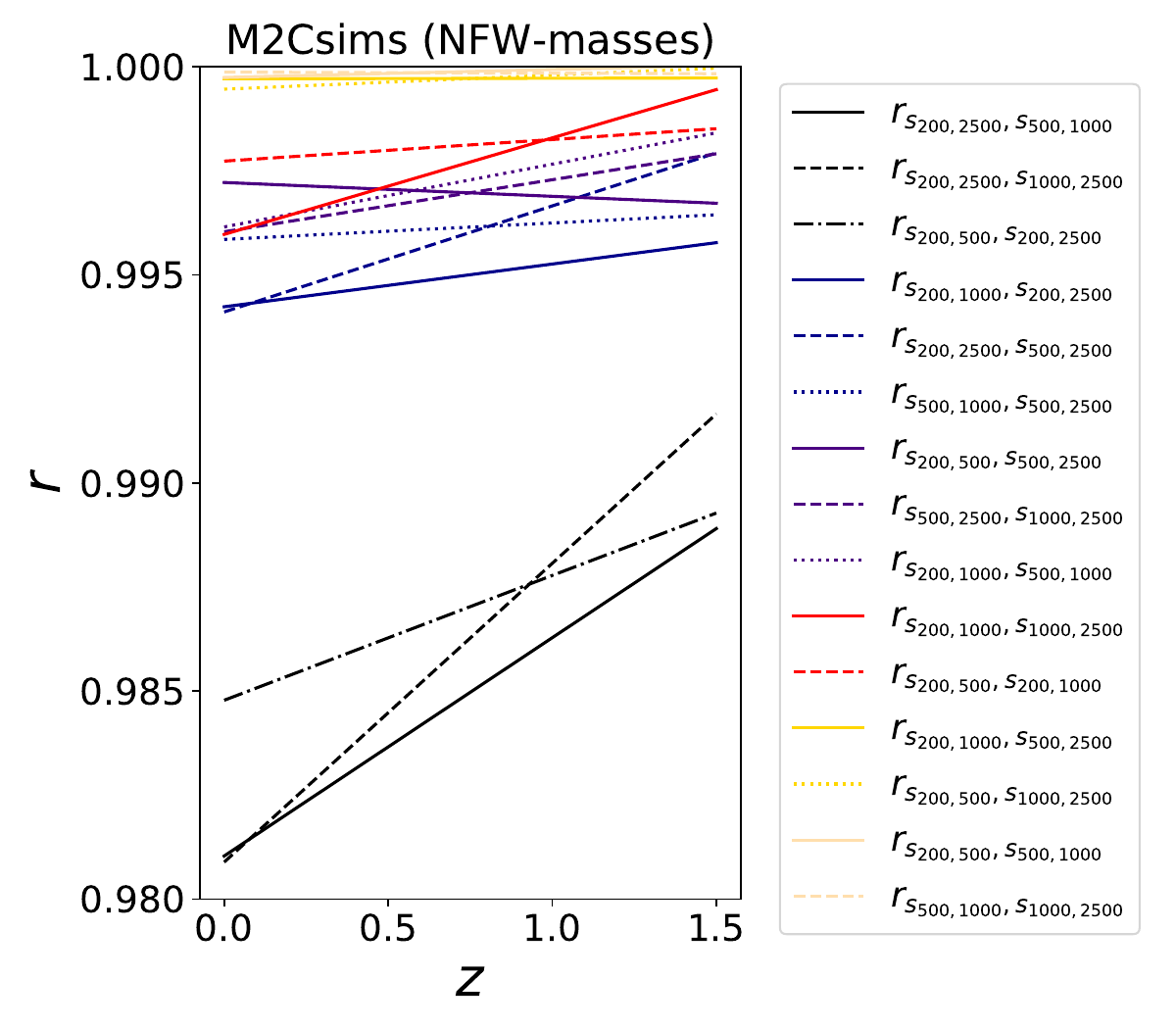}\
    \caption{Sparsity correlation coefficients obtained using the mass estimates from the best-fit NFW profiles of the M2Csims halo catalogues. As we can see, contrary to the correlations among sparsities estimated from the SO halo masses, assuming that the density profile of haloes is described by NFW profile artificially correlates sparsities to order $r\approx 1$, even those probing mass distribution in mass shells at large separations.}
    \label{fig:corr_nfw}
\end{figure}

Notice that the correlation coefficients from the Raygal catalogues slightly differ from those of the M2Csims ones. Again, this is a direct consequence of the differences between the simulated cosmological models. In particular, at any given redshift, the correlations from M2Csims are slightly larger than that from Raygal, which is consistent with the fact that the former has a larger $S_8$ value than the latter. Nevertheless, these differences are too small to have an impact on the cosmological parameter inference as we will discuss in Section~\ref{mcmc_analysis}.

In contrast, we would like to highlight the fact that if the density profile of haloes were exactly described by the Navarro-Frenk-White formula \citep[NFW,][]{NFW1997}, then all the information on the halo mass profile would be fully encoded in the values of the concentration parameter and the overall halo mass, given that the halo mass at any other overdensities can be derived from these two quantities. Furthermore, because of the one-to-one relation between halo sparsity and concentration for NFW haloes \citep[see][]{Balmes2014}, this would imply that a single sparsity estimate would carry all the information on the mass profile. Hence, if the density profile of haloes exactly were to follow the NFW model, one should find that different sparsities, even those probing distant mass shells, are highly correlated. One may argue that, given the fact that for each halo, the best-fit value of the concentration is a stochastic variable characterised by a scatter and a mean that varies with $M_{200c}$, then the correlation among the NFW inferred sparsities, e.g. $s_{200,\Delta_2}$ and $s_{\Delta_3,\Delta_4}$ (with $\Delta_2\ne \Delta_3\ne \Delta_4> 200$) may not be exactly one. However, because the functional form of the mass profile has to follow NFW, these should still be close to unity. This is indeed what we find when we compute the correlations among sparsities which have been computed using NFW inferred masses for each halo in the catalogues, as shown in Fig.~\ref{fig:corr_nfw}. More specifically, for each halo with a given mass $M_{200c}$ in the M2Csims catalogues, we have fit its density profile with the NFW function and deduced the corresponding best-fit NFW concentration parameter $c_{200c}$. Then, given the values of $M_{200c}$ and $c_{200c}$, we have calculated the NFW halo mass at $\Delta=500,1000$ and $2500$ (in units of the critical density) and computed the associated sparsities. Finally, we have estimated the correlation coefficients among the various sparsities using Eq.~(\ref{corr_coeff}). As we can see in Fig.~\ref{fig:corr_nfw}, the correlation coefficients among these NFW estimated sparsities are all close to unity. This is in sharp contrast with what we found from the analysis of the N-body halo masses shown in Fig.~\ref{fig:corr}. 

Indeed, the fact that differently from the NFW case, the correlations among different sparsities are not all clustered around $r=1$, but spread over a larger interval of values, as shown in Fig.~\ref{fig:corr}, is indicative of the fact that on average N-body haloes are not exactly described by the NFW formula. Moreover, it clearly shows that there is additional information about the halo mass profile which is not captured by the NFW profile but can be extracted using multiple sparsity measurements, thus potentially providing additional constraints on the cosmological parameters.

\section{Synthetic Data Analysis}\label{mcmc_analysis}
We seek to investigate the constraints on cosmological parameters that can be inferred from multiple sparsity measurements. To this end, we use the average sparsity estimates from the N-body halo catalogues as a synthetic dataset and perform a Markov Chain Monte Carlo likelihood analysis under different average sparsity error model assumptions. Our goal is twofold. On the one hand, we want to test to which extent the analysis recovers the fiducial cosmological parameters of the simulated cosmologies. On the other hand, we aim to study how the inferred parameter uncertainties vary for different sparsity configurations, uncertainties and fiducial cosmologies.

\subsection{Sparsity Configurations and Uncertainties}\label{spars_config}
We consider the following set of average sparsity combinations:
\begin{enumerate}
    \item[S1)] $\langle s_{200,2500}\rangle$;
    \item[S2)] $\langle s_{200,500}\rangle$, $\langle s_{200,2500}\rangle$;
    \item[S3)] $\langle s_{200,500}\rangle$, $\langle s_{200,2500}\rangle$, $\langle s_{500,2500}\rangle$;
    \item[S4)] $\langle s_{200,500}\rangle$, $\langle s_{200,1000}\rangle$, $\langle s_{200,2500}\rangle$, $\langle s_{500,2500}\rangle$; 
    \item[S5)]  $\langle s_{200,500}\rangle$, $\langle s_{200,1000}\rangle$, $\langle s_{200,2500}\rangle$, $\langle s_{500,1000}\rangle$, $\langle s_{500,2500}\rangle$;
    \item[S6)] $\langle s_{200,500}\rangle$, $\langle s_{200,1000}\rangle$, $\langle s_{200,2500}\rangle$, $\langle s_{500,1000}\rangle$, $\langle s_{500,2500}\rangle$, $\langle s_{1000,2500}\rangle$;
\end{enumerate}
where starting from the single sparsity $\langle s_{200,2500}\rangle$ we explore multiple sparsity configurations up to S6, which corresponds to the maximal number of sparsities $N_s$ that can be obtained from the estimation of halo masses at four different overdensities. 

In principle, for the configurations S1 to S5, we have a total of 62 possible sparsity configurations to study. Rather than a brute force investigation, for any $m<6$ number of sparsities, we have adopted a physically motivated strategy to explore among the various possibilities. This relies upon the observation that the cosmological differences among single average sparsity measurements $\langle s_{\Delta_1,\Delta_2}\rangle$ at a given redshift are maximised when the differences between $\Delta_1$ and $\Delta_2$ are the largest \citep{Balmes2014}. Hence, given the range of overdensities $\Delta$ we have considered, we set S1 to be the sparsity associated to the largest overdensity separation $\Delta^{\rm min}_1=200$ and $\Delta^{\rm max}_2=2500$. For the S2 configuration, we proceed by adding the average sparsity which probes the average mass distribution in a mass shell at an intermediate overdensity between $\Delta^{\rm min}_1$ and $\Delta^{\rm max}_2$. In our case we have chosen $\Delta=500$ and considered $\langle s_{200,500}\rangle$. This is done with the intent of investigating how the cosmological parameter constraints vary with the addition of information encoded within an intermediate mass shell with respect to the one already accounted by the previous sparsity configuration. For the configuration S3, we consider the average sparsity associated to the overdensities with the second largest separations among those considered at S2, that is $\langle s_{500,2500}\rangle$. Then, we proceed in a similar manner for S4 and S5.

For each of the sparsity combinations in the list, we consider two distinct synthetic datasets that consist of the average sparsity estimates at different redshifts from the Raygal and M2Csims catalogues, respectively. Given the larger number of redshift snapshots of the M2Csims simulations, this enable us to asses the impact of additional average sparsity estimates for a larger number of redshifts in the same redshift interval $0 \le z\le 1.5$. 

Here, we account for statistical uncertainties on average sparsity measurements and propagate the effect of systematic errors due to the mass function model uncertainties in predicting the redshift and cosmological model dependence of the average sparsity. In Section~\ref{forecast}, we extend the analysis of systematics and present the result of a forecast parameter inference analysis for realistic cluster survey configurations. 

Statistical errors on average sparsities are the consequence of the propagation of the uncertainties of cluster mass measurements\footnote{We neglect possible correlations among cluster mass determination, a choice which makes the assumed errors on the average sparsity only more conservative. In fact, given that the sparsity is a mass ratio, neglecting the correlations $r_{\Delta_1,\Delta_2}$ between the determination of the masses $M_{\Delta_1}$ and $M_{\Delta_2}$ is equivalent to overestimating the errors on the sparsity by a factor $\sim 1/\sqrt{1-r}$.}. Following \citet{Corasaniti2018}, we model the error on the average sparsity $\langle s_{\Delta_1,\Delta_2}\rangle$ at redshift $z$ as:
\begin{equation}\label{av_spars_err}
    \sigma_{\langle s(z)\rangle}= \frac{\langle s_{\Delta_1,\Delta_2}(z)\rangle}{\sqrt{N_{\rm cl}(z)}}  \sqrt{e^2_{M_{\Delta_1}}+e^2_{M_{\Delta_2}}},
\end{equation}
where $e_{M_{\Delta_1}}$ and $e_{M_{\Delta_2}}$ are the fractional error on the mass measurements at overdensities $\Delta_1$ and $\Delta_2$ respectively, and $N_{\rm cl}(z)$ is the number of clusters in the bin centred at redshift $z$.

We focus on a simplified configuration and consider two distinct cases for the statistical errors: $\sigma_{\langle s(z)\rangle}=0.3$ and $0.1$. The former is a rather conservative choice, corresponding to having $\sim 50$ clusters per redshift bin with $30\%$ fractional mass measurement errors, while the latter is a more optimistic assumption corresponding to having $\sim 100$ clusters per redshift bin with $10\%$ precision on the estimated masses. 

We treat the discrepancies between the average sparsity predictions and the N-body found in Section~\ref{average_spars} and Appendix \ref{app_a} as an intrinsic systematic error $\sigma^{\rm sys}_{s_{\Delta_1,\Delta_2}}(z)$ on the average sparsity $\langle s_{\Delta_1,\Delta_2}\rangle$ obtained by solving Eq.~(\ref{spars_mf}).

\subsection{Priors \& Likelihood}\label{priorslike}
We specifically focus our analysis on $\Omega_m$ and $\sigma_8$, the cosmological parameters to which the sparsity is most sensitive, while setting $h$, $\Omega_b$ and $n_s$ to their fiducial values. We assume uniform priors on $\Omega_m\sim U(0.1,0.5)$ and $\sigma_8\sim U(0.2,1.2)$.

We perform a MCMC sampling of the log-likelihood function:
\begin{equation}\label{loglike}
    -2 \ln{\mathcal{L}}=\sum_{i,j=1}^{m}\sum_{k=1}^{N_z}\Delta{s_i(z_k)}\cdot C_{s_i,s_j}^{-1}(z_k)\cdot \Delta{s_j(z_k)},
\end{equation}
with $m\le N_s$ the number of sparsity configurations considered, $N_z$ the number of redshift bins and 
\begin{equation}\nonumber
    \Delta{s_i(z_k)}=\langle s^{\rm mf}_i(z_k)\rangle(1+\tilde{Y}^{s_i}_{z_k})-\langle s_i(z_k) \rangle,
\end{equation}
where $\langle s_i(z_k) \rangle$ is the synthetic data point at the $k$-th redshift bin $z_k$ for the $i$-th configuration of overdensities, while $\langle s^{\rm mf}_i(z_k)\rangle$ is the average sparsity predicted by the mass function model Eq.~(\ref{spars_mf}) with $\tilde{Y}^{s_i}_{z_k}\sim N(0,\sigma^{\rm sys}_{s_i}(z_k))$ being a Gaussian random variable which we marginalise over, characterised by zero mean and standard deviation $\sigma^{\rm sys}_{s_i}(z)$. The latter being the sum of the intrinsic scatter with respect to the N-body average sparsities discussed in Section~\ref{average_spars} and Appendix~\ref{app_a}. The covariance matrix reads as:
\begin{equation}\label{cov_av_spars}
    C_{s_i,s_j}(z_k)=\sigma_{\langle s(z_k) \rangle}^2 r_{s_i,s_j}(z_k),
\end{equation}
where $r_{s_i,s_j}(z_k)$ is the correlation matrix at redshift $z_k$, which we have previously computed using the Raygal and M2Csims catalogues in Section~\ref{av_corr} and $\sigma_{\langle s(z_k) \rangle}$ is the statistical uncertainty on the average sparsity estimates.

\begin{table*}
\centering
    \caption{Marginalised $1\sigma$ errors on $\Omega_m$, $\sigma_8$ and $S_8$ from the MCMC likelihood analysis of the Raygal and M2Csims synthetic data inferred assuming average sparsity errors of $\sigma_{\langle s(z)\rangle}=0.3$ and $0.1$ respectively, for the various sparsity configurations S1-S6. As we can see, the constraints on the cosmological parameters improve for increasing number of sparsity configurations, reaching a minimum for S4.}
    \begin{tabular}{c | ccc | ccc | ccc | ccc}
    \hline
    & \multicolumn{3}{c}{Raygal $(\sigma_{\langle s(z)\rangle}=0.3)$}   &
    \multicolumn{3}{c}{Raygal $(\sigma_{\langle s(z)\rangle}=0.1)$} & \multicolumn{3}{c}{M2Csims $(\sigma_{\langle s(z)\rangle}=0.3)$} &
    \multicolumn{3}{c}{M2Csims $(\sigma_{\langle s(z)\rangle}=0.1)$}\\
    \hline
    Configuration & $\sigma_{\Omega_m}$ & $\sigma_{\sigma_8}$ & $\sigma_{S_8}$ & $\sigma_{\Omega_m}$ & $\sigma_{\sigma_8}$ & $\sigma_{S_8}$ &
    $\sigma_{\Omega_m}$ & $\sigma_{\sigma_8}$ & $\sigma_{S_8}$ &
    $\sigma_{\Omega_m}$ & $\sigma_{\sigma_8}$ & $\sigma_{S_8}$ \\
    \hline
    S1 & $0.039$ & $0.124$ & $0.086$ & 
    $0.019$ & $0.061$ & $0.038$ & 
    $0.068$ & $0.100$ & $0.042$ &
    $0.043$ & $0.060$ & $0.018$\\
    S2 & $0.037$ & $0.109$ & $0.072$ &
    $0.016$ & $0.049$ & $0.030$ &
    $0.061$ & $0.086$ & $0.037$ &
    $0.034$ & $0.047$ & $0.015$ \\
    S3 & $0.007$ & $0.027$ & $0.019$ &
    $0.003$ & $0.012$ & $0.008$ &
    $0.016$ & $0.025$ & $0.008$ &
    $0.007$ & $0.010$ & $0.004$ \\
    S4 & $0.002$ & $0.008$ & $0.005$ &
    $0.002$ & $0.010$ & $0.006$ &
    $0.007$ & $0.009$ & $0.002$ &
    $0.006$ & $0.008$ & $0.001$ \\
    S5 & $0.005$ & $0.020$ & $0.015$ &
    $0.002$ & $0.011$ & $0.007$ &
    $0.011$ & $0.017$ & $0.007$ & 
    $0.006$ & $0.007$ & $0.003$ \\
    S6 & $0.005$ & $0.020$ & $0.015$ &
    $0.002$ & $0.011$ & $0.007$ &
    $0.011$ & $0.017$ & $0.007$ &
    $0.006$ & $0.007$ & $0.003$ \\
    \hline
    \end{tabular}\label{table_cosmopar}
\end{table*}

\subsection{Results}
We use the MCMC chains to infer marginal constraints on $\Omega_m$, $\sigma_8$ and $S_8$. For this purpose, we have implemented a Metropolis-Hastings algorithm and tested the convergence of the chains with the Gelman-Rubin diagnostics \citep[see][for a review]{2020AnRSA...7..387R}. We have analysed the chains using the publicly available package \texttt{GetDist}\footnote{\href{https://getdist.readthedocs.io/}{https://getdist.readthedocs.io/}} \citep{2019arXiv191013970L}. The results are summarised in Table~\ref{table_cosmopar} where we quote the marginalised $1\sigma$ errors on the parameters as obtained from the various cases we have considered. We plot the corresponding $1$ and $2\sigma$ credibility contours in the $\Omega_m-\sigma_8$ from the Raygal synthetic data analysis in Fig.~\ref{fig:2D_raygal} and for the M2Csims dataset in Fig.~\ref{fig:2D_m2csims}.

\subsubsection{Raygal Synthetic Dataset}
\begin{figure*}
    \centering
    \includegraphics[width = 0.48\linewidth]{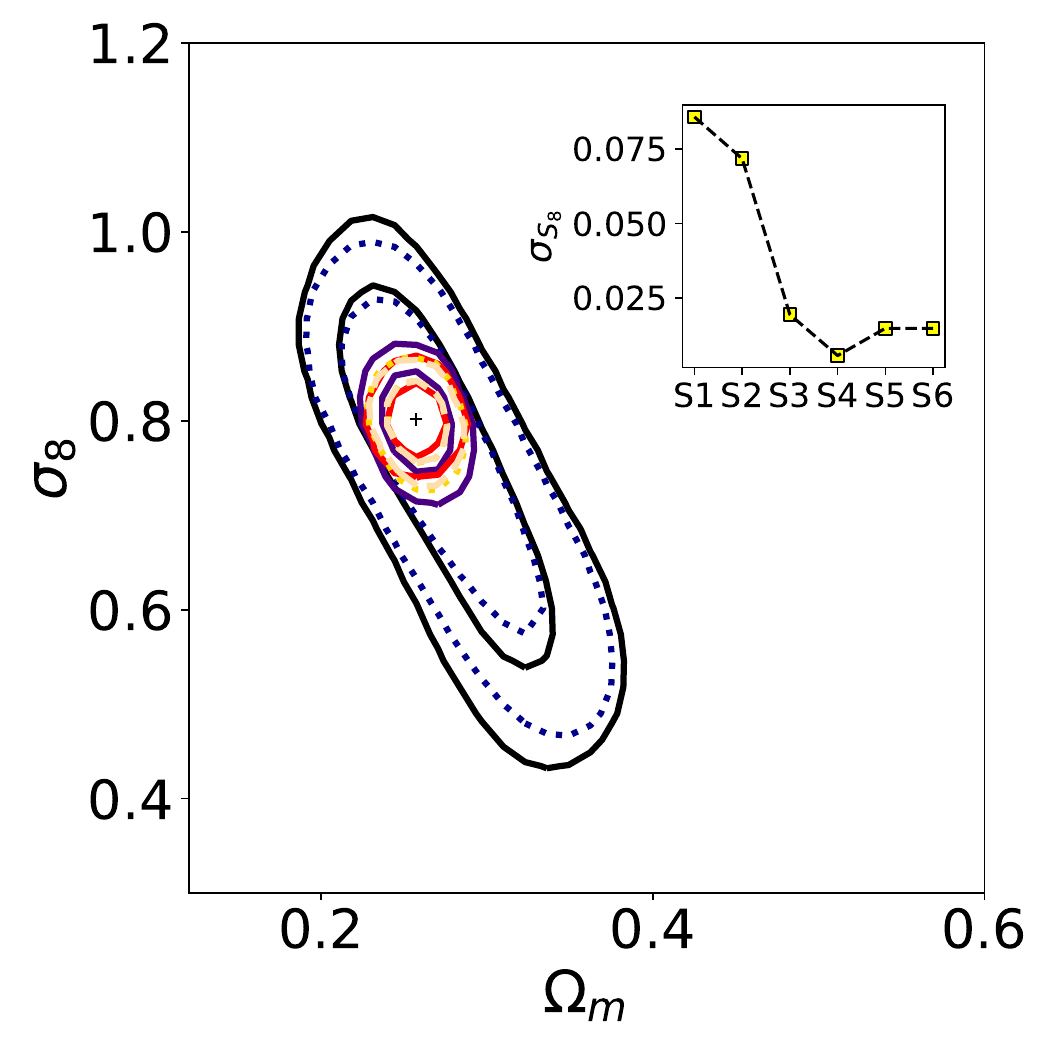}\
    \includegraphics[width = 0.48\linewidth]{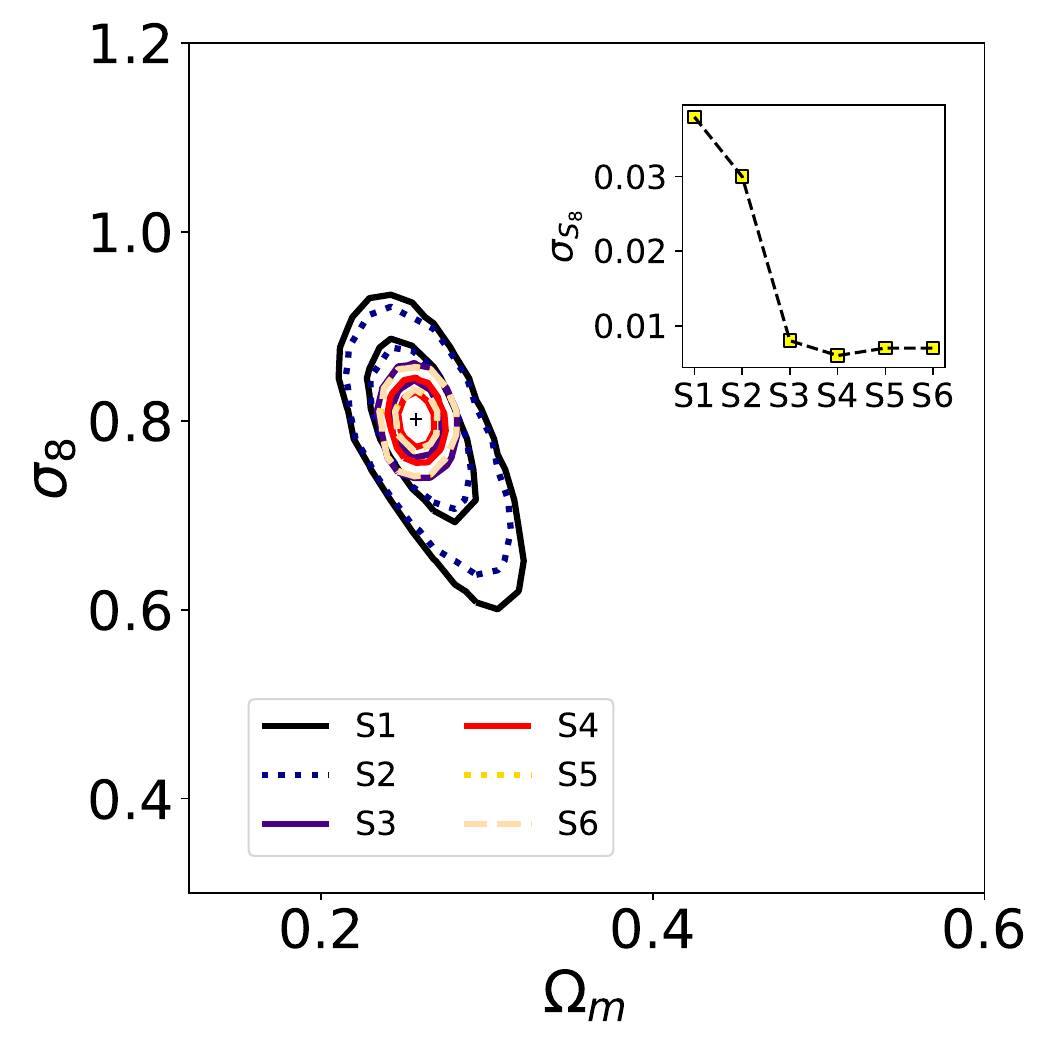}
    \caption{$1$ and $2\sigma$ credibility contours in the $\Omega_m-\sigma_8$ plane for the Raygal synthetic datasets. The different lines correspond to the sparsity configurations S1-S6 described in Section~\ref{spars_config} assuming statistical errors on the average sparsity estimates of $\sigma_{\langle s(z\rangle)}=0.3$ (left panel) and $\sigma_{\langle s(z\rangle)}=0.1$ (right panel). The cross corresponds to the cosmological parameter values of the fiducial Raygal cosmology. As the best-fit parameters for the different sparsity configurations coincide with those of the fiducial model, we do not show them in the plot to avoid cluttering. The inset plot shows the $1\sigma$ error on $S_8$ as a function of the number of average sparsity configurations S1-S6. As we can see the uncertainties saturate beyond S4.}
    \label{fig:2D_raygal}
\end{figure*}

Firstly, we find that for all sparsity configurations S1-S6, the best-fit parameters of the MCMC likelihood analysis coincide with the values of the Raygal fiducial cosmology. Unsurprisingly, the constraints on $\Omega_m$ and $\sigma_8$ (and consequently $S_8$) quoted in Table~\ref{table_cosmopar} show that, for a given sparsity configuration, the inferred parameter errors obtained assuming statistical errors of $\sigma_{\langle s(z)\rangle}=0.1$ are systematically smaller than those obtained for $\sigma_{\langle s(z)\rangle}=0.3$. This can also be seen by the different size of the $1$ and $2\sigma$ credibility contours shown in Fig.~\ref{fig:2D_raygal}. Quite interestingly, we notice that, in both cases, the contours shrink from S1 to S4, thus indicating that using additional sparsity measurements does improve the cosmological constraints. On the other hand, we can see that, in the case of a greater number of sparsity configurations S5 and S6, the contours do not shrink further. Quite the opposite, from the marginalised error values quoted in Table~\ref{table_cosmopar}, we find that the constraints on the cosmological parameters slightly degrade. This is most likely due to the fact that the additional sparsities considered in S5 and S6, namely $\langle s_{500,1000}\rangle$ and $\langle s_{1000,2500}\rangle$, do not vary significantly with the cosmological parameters. As we have seen in Section~\ref{average_spars}, these have variation of the order of percent level around the fiducial cosmology, which is of the same order of the accuracy of the cosmological model predictions given by Eq.~\ref{spars_mf} using the numerical ST calibrated mass functions. This trend can be better seen in the inset plots, where we show the marginalised $1\sigma$ error on $S_8$ as function of the sparsity configurations considered. As we can see, $\sigma_{S_8}$ diminishes as function of the number of sparsity configuration considered reaching a minimum value for S4, while increasing for S4 and S5 a possible consequence of the $\Omega_m-\sigma_8$ degeneracy. Indeed, taking as figure-of-merit the values of the area within the $1\sigma$ credibility contours in the $\Omega_m-\sigma_8$ plane highlights better the saturation of the constraints on the cosmological parameters beyond S4. As we can infer from the values quoted in Table~\ref{area_raygal}, the area diminishes from S1 to S4 and then remains constant.

\begin{table}
\centering
\caption{Area within the $1\sigma$ credibility contours shown in Fig.~\ref{fig:2D_raygal} for the various sparsity configuration in the case of the Raygal data analysis with $(\sigma_{\langle s(z)\rangle}=0.3)$ and $0.1$ statistical uncertainties. We may notice that the area diminishes for increasing number of sparsity configurations and saturates at S4.}
\begin{tabular}{ ccc }
\hline 
 & Raygal Analysis & \\
\hline
Configuration & $A_{1\sigma}$ $(\sigma_{\langle s(z)\rangle}=0.3)$ & $A_{1\sigma}$ $(\sigma_{\langle s(z)\rangle}=0.1)$\\
\hline
S1  & $0.059$ & $0.021$ \\
S2  & $0.048$ & $0.017$ \\
S3  & $0.009$ & $0.005$ \\
S4  & $0.006$ & $0.004$ \\
S5  & $0.006$ & $0.004$ \\
S6  & $0.006$ & $0.004$ \\
\hline
\end{tabular}\label{area_raygal}
\end{table}

Quantitatively, we find that in the case of $\sigma_{\langle s(z)\rangle}=0.3$, the marginalised $1\sigma$ error on $\Omega_m$ improves by approximately a factor of 20 from S1 to S4; similarly, the uncertainties on $\sigma_8$ improves by a factor $\sim 15$. Even, considering three sparsity measurements, such as in the S3 configuration, leads to an improvement of a factor $\sim 5$ on $\sigma_{\Omega_m}$ and a factor $\sim 4$ on $\sigma_{\sigma_8}$. In the case of $\sigma_{\langle s(z)\rangle}=0.1$, we find that the $\sigma_{\Omega_m}$ reduces by a factor of $\sim 9$ from S1 to S4 and $\sigma_{\sigma_8}$ by a factor of $\sim 8$. On the other hand, it is worth noticing that, for S4, the constraints do not significantly improve when reducing the statistical errors on the average sparsities by a factor of $3$ (i.e. from $\sigma_{\langle s(z)\rangle}=0.3$ to $0.1$). This suggests that the use of fours sparsity measurements can mitigate the need for improved mass measurements.

\subsubsection{M2Csims Synthetic Dataset}
\begin{figure*}
    \centering
    \includegraphics[width = 0.48\linewidth]{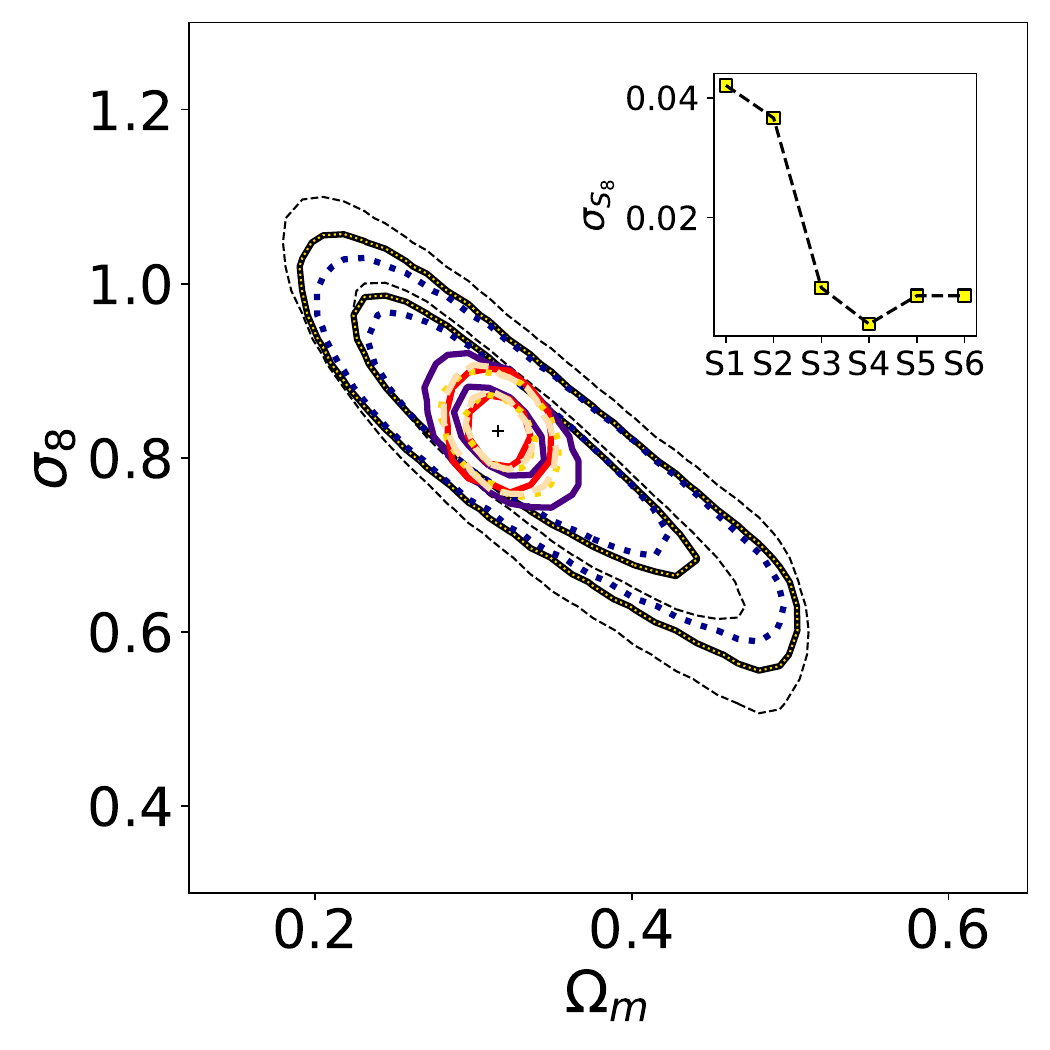}\
    \includegraphics[width = 0.48\linewidth]{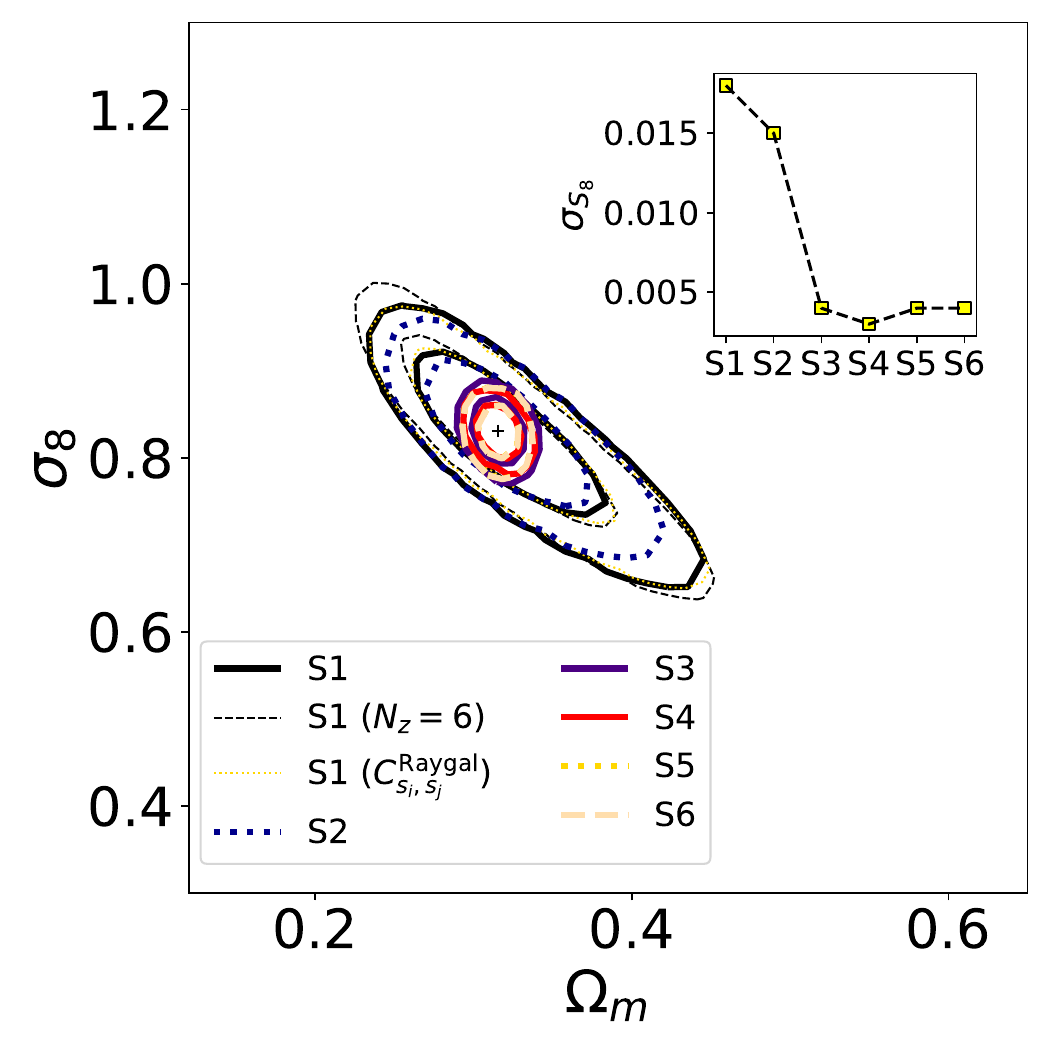}
    \caption{As in Fig.~\ref{fig:2D_raygal} but for the M2Csims case. In addition to the sparsity combinations S1-S6, we also show the $1$ and $2\sigma$ contours for the S1 case with $N_z=6$ redshift bins in the redshift interval $0<z<1.5$ (thin black dashed lines) rather than the nominal $N_z=11$ (black solid line), and with covariance from the Raygal simulation (yellow dotted line).}
    \label{fig:2D_m2csims}
\end{figure*}

The likelihood analysis of the MC2sims dataset shows trends that are similar to those we have found using the Raygal dataset. Again, we have that, in all the cases we considered, the best-fit parameters coincide with the values of the MC2sims fiducial cosmology. Moreover, as it can be noticed from the values quoted in Table~\ref{table_cosmopar} for $\sigma_{\langle s(z)\rangle}=0.3$ and $0.1$, also in these cases we find that the uncertainties on $\Omega_m$, $\sigma_8$ (and $S_8$) decrease to a minimum value as the number of sparsities increases from S1 to S4, while they slightly increase for S5 and S6 configurations. This can also be seen in Fig.~\ref{fig:2D_m2csims}, where the credibility contours shrink from S1 to S4 as we have already found in the Raygal case. The inset plot shows $\sigma_{S_8}$ as function of the number of sparsity configurations, which reaches a minimum value for S4 and slightly increase for S5 and S6. Again, we can better appreciate the saturation of the cosmological parameter constraints at S4 from the values of the area enclosed within the $1\sigma$ credibility contours quoted in Table~\ref{area_m2csims}.

\begin{table}
\centering
\caption{As in Table~\ref{area_raygal} for the M2Csims data analysis.}
\begin{tabular}{ ccc }
\hline 
 & M2Csims Analysis & \\
\hline
Configuration & $A_{1\sigma}$ $(\sigma_{\langle s(z)\rangle}=0.3)$ & $A_{1\sigma}$ $(\sigma_{\langle s(z)\rangle}=0.1)$\\
\hline
S1  & $0.058$ & $0.030$ \\
S2  & $0.050$ & $0.024$ \\
S3  & $0.011$ & $0.004$ \\
S4  & $0.007$ & $0.003$ \\
S5  & $0.007$ & $0.003$ \\
S6  & $0.007$ & $0.003$ \\
\hline
\end{tabular}\label{area_m2csims}
\end{table}

Quantitatively, from the values quoted in Table~\ref{table_cosmopar}, we find an improvement of a factor of $\sim 10$ on $\sigma_{\Omega_m}$ and $\sigma_{\sigma_8}$ for $\sigma_{\langle s(z)\rangle}=0.3$, and a factor of $\sim 7$ for $\sigma_{\langle s(z)\rangle}=0.1$. 

Notice that in addition to the contours from the analysis of the configurations S1-S6, in Fig.~\ref{fig:2D_m2csims}, we also plot the results of two additional cases we have investigated for the S1 configuration. In particular, we have performed an analysis of the M2Csims synthetic average sparsity data using the covariance matrix from the Raygal simulation such as to evaluate the impact of the cosmological dependence of the covariance on the cosmological parameter constraints. For this purpose, we have evaluated the covariance at the M2Csims redshift bins using the parametrisation of the Raygal sparsity correlation coefficients given in Appendix~\ref{app_b}. We have also performed an analysis of the M2Csims synthetics dataset limited to $N_z=6$ redshift bins at $z=0.00,0.49,0.61,1.00,1.27$ and $1.50$ (as in the Raygal case) to evaluate the impact of additional redshift bins on the cosmological parameter inference. In the former case, we find that there is no effect of using the Raygal covariance for the analysis of the M2Csims data, which suggests that the cosmological dependence of the covariance discussed in Section~\ref{av_corr} is too small to have an impact on the cosmological parameter inference for the level of average sparsity uncertainty we have assumed. In the latter case, we do find that increasing the number of redshift bins improves the constraints on the parameters, though not significantly when compared to the effect of using multiple sparsity measurements. As an example, for the case $\sigma_{\langle s(z)\rangle}=0.3$ with $N_z=6$, we find $\sigma_{\Omega_m}=0.074$ and $\sigma_{\sigma_8}=0.118$, while in the case with $N_z=11$, we have $\sigma_{\Omega_m}=0.068$ and $\sigma_{\sigma_8}=0.100$. Similarly, for the case with $\sigma_{\langle s(z)\rangle}=0.1$ and $N_z=6$, we have $\sigma_{\Omega_m}=0.047$ and $\sigma_{\sigma_8}=0.072$, while in the case with $N_z=11$, we have $\sigma_{\Omega_m}=0.043$ and $\sigma_{\sigma_8}=0.060$.

Overall, comparing the results from the Raygal analysis and those obtained from the M2Csims, we find that the inferred parameter constraints do depend on the underlying fiducial cosmology. As summarised by the $1\sigma$ errors on $S_8$, we have that for a given sparsity configuration and given level of statistical uncertainty on the average sparsity measurements, the value of $\sigma_{S_8}$ is systematically smaller in the M2Csims case than in the Raygal case by approximately a factor of $2$. Such dependence on the fiducial cosmology of forecast parameter error analysis is not new \citep[e.g. we refer the readers to the Appendix B of][for a detailed discussion]{2006MNRAS.369.1725M}. It simply reflects the amplitude of the variation of the observable (the average sparsity in our case) across the cosmological parameter space relative to the amplitude of the observational uncertainties at the observed data points. This justifies the need for parameter forecast studies performed under different model assumptions. 


\section{CHEX-MATE Clusters Forecast Analysis}\label{forecast}
We forecast cosmological parameter constraints from multiple average sparsity measurements for a realistic galaxy cluster data sample. We specifically focus on cluster mass measurements as expected from the CHEX-MATE project \citep{2021A&A...650A.104C}, which consists of a sample of 118 clusters from the Planck-SZ catalogue in the redshift range $0< z<0.6$. These are the targets of a dedicated X-ray observing program on the \textit{XMM}-Newton satellite, which is expected to provide accurate measurements of the cluster mass distributions and gas properties. For each cluster in the sample, mass estimates at different overdensities will be obtained under the hydrostatic equilibrium (HE) hypothesis. Similarly to the study presented in Section~\ref{mcmc_analysis}, we perform a likelihood MCMC analysis of a synthetic dataset with characteristics and mass measurement errors expected from the CHEX-MATE sample to infer constraints on $\Omega_m$ and $\sigma_8$. In the following, we set the fiducial cosmological model to the flat $\Lambda$CDM best-fit to the Planck-2015 data \citep{2016A&A...594A..13P}.

\begin{figure}
    \centering
    \includegraphics[width = 0.9\linewidth]{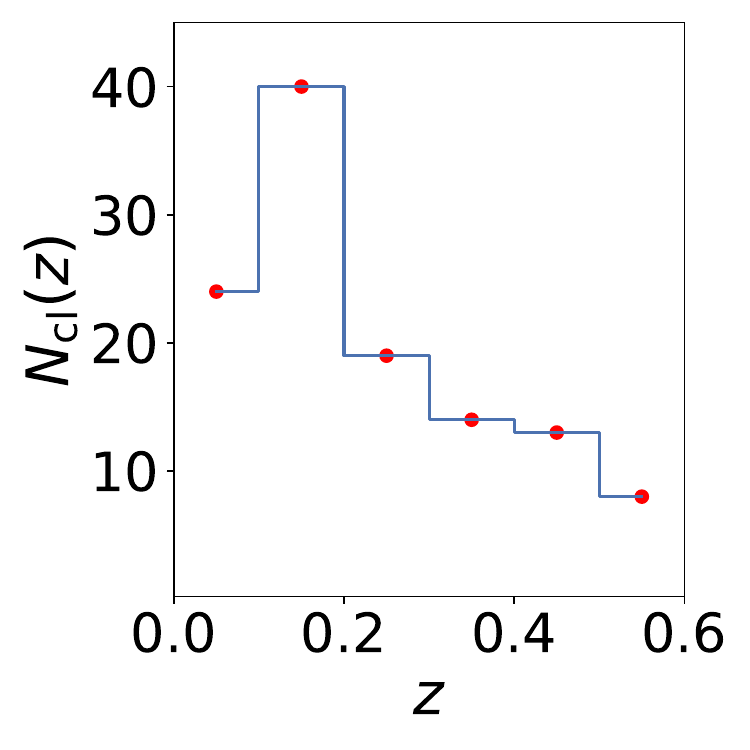}
    \caption{CHEX-MATE binned cluster counts in equally-spaced redshift bins of size $\Delta{z}=0.1$ in the range $0<z<0.6$.}
    \label{fig:nz}
\end{figure}

\begin{figure}
    \centering
    \includegraphics[width = 0.9\linewidth]{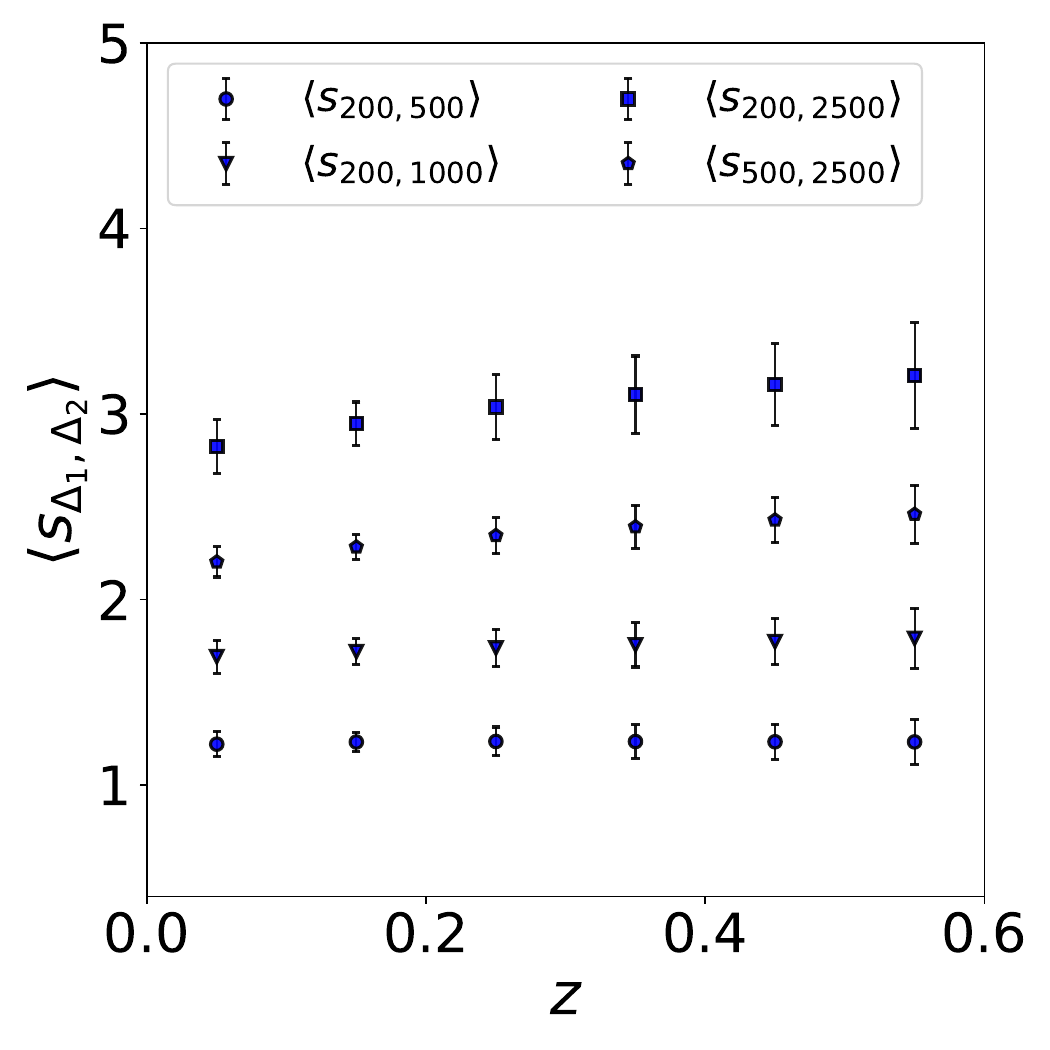}
    \caption{Synthetic average sparsity data $\langle s_{200,500}\rangle$ (circles), $\langle s_{200,1000}\rangle$ (triangles), $\langle s_{200,2500}\rangle$ (squares) and $\langle s_{500,2500}\rangle$ (pentagons). The error bars indicate the amplitude of statistical errors due to the propagation of mass measurement uncertainties.}
    \label{fig:spars_chex_syn}
\end{figure}

 In order to build a synthetic dataset of average sparsity measurements that are consistent with the characteristics of the CHEX-MATE sample, we first bin the CHEX-MATE clusters in equally spaced redshift bins of size $\Delta{z}=0.1$, the corresponding number counts $N(z)$ are shown in Fig.~\ref{fig:nz}. Then, we generate a sample of synthetic average sparsity data $\langle s_{200,500}\rangle$, $\langle s_{200,1000}\rangle$, $\langle s_{200,2500}\rangle$ and $\langle s_{500,2500}\rangle$ by solving Eq.~(\ref{spars_mf}) at the central redshift of the different bins using the M2Csims mass function parametrisations discussed in Appendix~\ref{app_a}. We also estimate the average sparsity errors using Eq.~(\ref{av_spars_err}), where we assume cluster mass uncertainties expected from the analysis of the CHEX-MATE observations. In particular, thanks to the observational strategy adopted in CHEX-MATE, homogenous exposures with \textit{XMM-Newton} for the entire sample will guarantee to reach a relative error of about $15\%$ on the hydrostatic masses measured at $\Delta=500$. Hence, by interpolating and scaling the relative errors on hydrostatic masses obtained at different overdensities in the X-COP project \citep{2019A&A...621A..39E}, we can reasonably assume fractional mass errors of $e_{M_{\Delta}}=0.23,0.15,0.11$ and $0.10$ at $\Delta=200,500,1000$ and $2500$, respectively. The synthetic datasets are shown in Fig.~\ref{fig:spars_chex_syn}.
 
 We consider two distinct cases: single average sparsity measurements $\langle s_{200,2500}\rangle$ (S1); four average sparsity measurements $\langle s_{200,500}\rangle$, $\langle s_{200,1000}\rangle$, $\langle s_{200,2500}\rangle$ and $\langle s_{500,2500}\rangle$ (S4). In the latter case, we evaluate the covariance matrix using Eq.~(\ref{cov_av_spars}), where we estimate the correlation coefficients for the different average sparsities at the different redshifts using the linear regression obtained from the analysis of the M2Csims halo catalogues and presented in Appendix~\ref{app_b}.
  
 We assume the log-likelihood function as given by Eq.~(\ref{loglike}). Similarly to the analysis presented in Section~\ref{mcmc_analysis}, we propagate the effect of systematic uncertainties by marginalising over the Gaussian random variable with zero mean and standard deviation corresponding to the sum of all systematic errors we account for. Here, in addition to the intrinsic scatters due to our data model, we also propagate the impact of mass biases on the average sparsity caused by the presence of baryons.

We infer constraints for three different error configurations: 
\begin{enumerate}
    \item[(a)] statistical errors due to the propagation of mass-measurement uncertainties as estimated by Eq.~(\ref{av_spars_err}) in combination with the intrinsic data model errors; 
    \item[(b)] statistical errors in combination with the intrinsic systematic errors of our data model and the systematic uncertainties due to the effects of hydrostatic mass bias on sparsity estimates;
    \item[(c)] statistical errors in combination with intrinsic systematic data model errors and systematic uncertainties due to the effects of baryons on sparsity estimates based on dark matter only masses.
\end{enumerate}
In cases (b) and (c), we estimate the impact of mass biases on the average sparsity by evaluating the percentage bias shift: $\Delta{b}_{\Delta_1,\Delta_2}=\Delta{s_{\Delta_1,\Delta_2}}/\langle s_{\Delta_1,\Delta_2}\rangle$.

We assume the percentage bias shifts due to HE mass bias estimated in \citet{2022MNRAS.513.4951R}, which have been obtained from the analysis of N-body/hydro simulations of galaxy clusters from \citet{2016ApJ...827..112B}. We quote these systematic bias shifts in Tab.~\ref{tab_bias}. Instead, we evaluate the impact of baryons on the average sparsity estimates from dark matter only masses using the results of the mass biases found in \citet{2014MNRAS.442.2641V} from the analysis of a combination of the OverWhelmingly Large Simulations (OWLS) \citep{2010MNRAS.402.1536S} and cosmo-OWLS \citep[][]{2014MNRAS.441.1270L} for the feedback model AGN 8.0 that reproduce the observed X-ray profiles of clusters \citep{2014MNRAS.441.1270L}. The corresponding percentage bias shifts on different sparsity estimates have been estimated in \citet{Corasaniti2018} as a function of cluster mass $M_{200c}>10^{13}$ M$_{\odot}$ h$^{-1}$ (see their Fig.~7). Here, we conservatively assume the largest absolute values from \citet{Corasaniti2018}, which we quote in Tab.~\ref{tab_bias}. 
 
\begin{table}
    \centering
    \caption{Percentage bias shift of the average sparsities due to the HE mass bias (first row) and the impact of baryons (second row).}
    \begin{tabular}{ccccc}
    \hline
 &$\Delta{b}_{200,500}$ & $\Delta{b}_{200,1000}$ & $\Delta{b}_{200,2500}$ & $\Delta{b}_{500,2500}$  \\
 \hline
 HE mass bias &  $0.03$ & $0.02$ & $0.03$ & $0.04$ \\
 Baryon mass bias & $0.04$ & $0.10$ & $0.15$ & $0.10$\\
    \hline
    \end{tabular}
    \label{tab_bias}
\end{table}

\begin{figure}
    \centering
    \includegraphics[width = 0.9\linewidth]{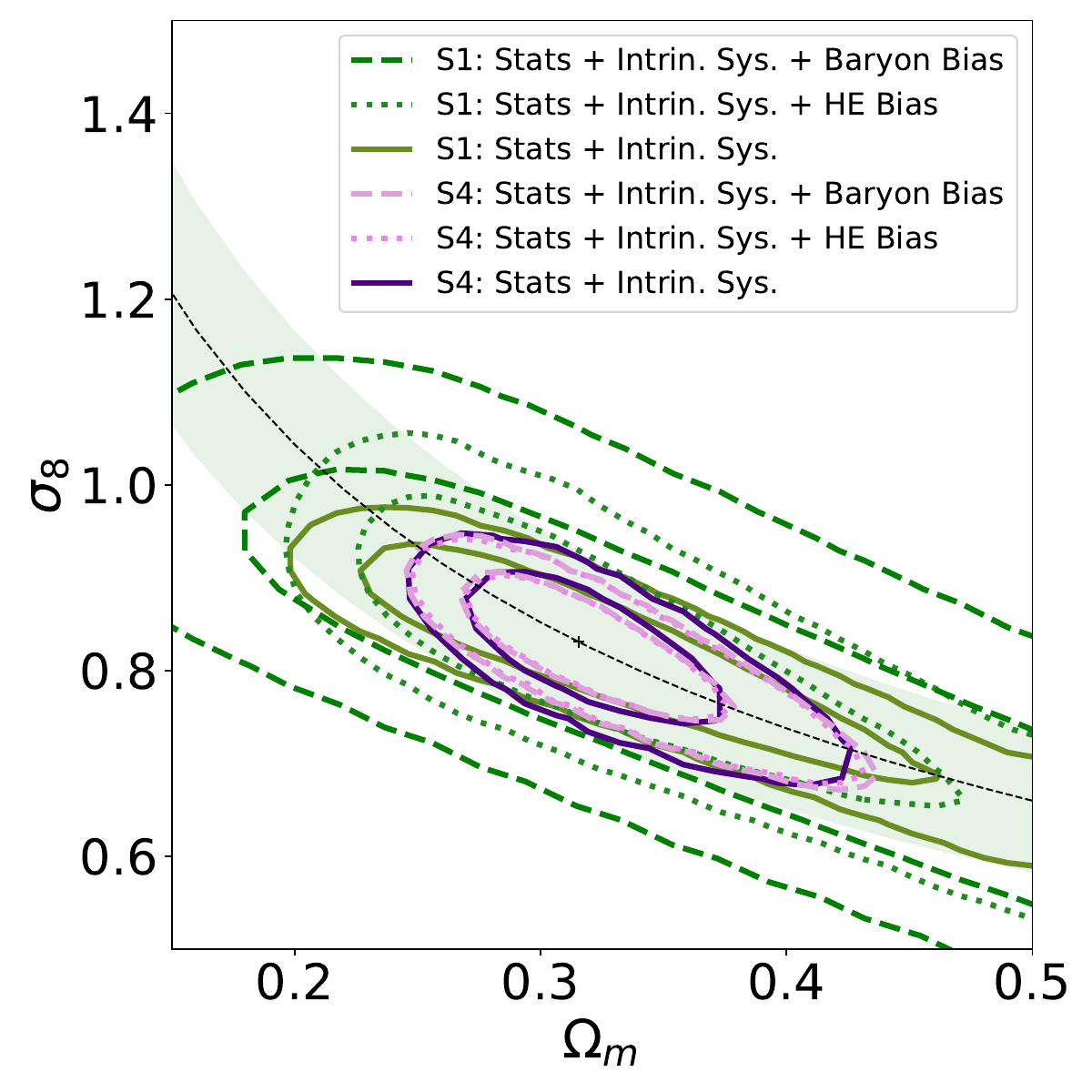}
    \caption{$1$ and $2\sigma$ credibility regions in the $\Omega_m-\sigma_8$ plane from the analysis of the synthetic dataset with different error assumptions for the S1 and S4 cases respectively. The best-fit values of $\Omega_m$ and $\sigma_8$ from the different parameter inferences coincide with the values of the fiducial cosmology marked by the cross-point. The dashed line and the green shaded area corresponds to curves of constant $S_8=0.852\pm 0.104$ values, that is the mean and standard deviation of $S_8$ from the MCMC chains of S1 case with intrinsic systematic errors and baryon bias. We can see that using multiple sparsity estimates breaks the $S_8$ degeneracy.}
    \label{fig:2D_chex}
\end{figure}

We assume priors and evaluate the likelihood as specified in Section~\ref{priorslike}. From the MCMC chains, we derive the marginal constraints on $\Omega_m$, $\sigma_8$ and $S_8$ for the different sparsity configurations and error assumptions. These are quoted in Tab.~\ref{tab:chexmate_res}, while in Fig.~\ref{fig:2D_chex}, we plot the corresponding $1$ and $2\sigma$ credibility regions in the $\Omega_m-\sigma_8$ plane. 

First of all, we find that in all the cases the best-fit model parameters coincide with those of the fiducial cosmological model, shown as a cross in Fig.~\ref{fig:2D_chex}. In the S1 case, we may notice that the inclusion of the systematic errors due to the HE or baryon biases only allow to infer an upper bound on $\sigma_8$ and a lower bound on $\Omega_m$. This is because, for the assumed errors, a single sparsity measurements over the range of redshift considered only constrains the degenerate parameter combination given by $S_8$. In particular, we find $\sigma_{S_8}=0.06$ for the HE bias and $\sigma_{S_8}=0.10$ for the baryon bias. The latter case is shown in Fig.~\ref{fig:2D_chex} as the green shaded region around the curve of constant best-fit value of $\hat{S}_8=0.85$. Such a result is consistent with the constraint obtained in \citet{Corasaniti2018} from the analysis of $s_{500,1000}$ of a sample of $\sim 100$ X-ray clusters. 

\begin{table*}\label{tab:chexmate_res}
    \centering
    \caption{Marginalised $1\sigma$ errors on $\Omega_m$, $\sigma_8$ and $S_8$ for the different sparsity configurations and error assumptions. In the last column, we quote the values of the area under the $1\sigma$ credibility countour. Notice that, in the S1 case, the propagation of systematic uncertainties due to the HE bias or the effect of baryons only allow to infer an upper bound on $\sigma_8$ and a lower bound on $\Omega_m$.}
    \begin{tabular}{ccccc}
    \hline
    & $\sigma_{\Omega_m}$ & $\sigma_{\sigma_8}$ & $\sigma_{S_8}$ & $A_{1\sigma}$ \\
    \hline
    S1 (Stats + Intrinsic Sys.) & $0.07$ & $0.08$ & $0.04$ & $0.014$\\
    S1 (Stats + Intrinsic Sys. + HE Bias) & $-$ & $-$ & $0.06$ & $-$\\ 
    S1 (Stats + Intrinsic Sys. + Baryon Bias) & $-$ & $-$ & $0.10$ & $-$ \\
    S4 (Stats + Intrinsic Sys.) & $0.04$ & $0.05$ & $0.02$ & $0.021$\\
    S4 (Stats + Intrinsic Sys. + HE Bias) & $0.04$ & $0.05$ & $0.02$ & $0.022$\\
    S4 (Stats + Intrinsic Sys. + Baryon Bias) & $0.04$ & $0.05$ & $0.02$ & $0.024$\\
    \hline
    \end{tabular}
    \end{table*}

In the S4 case, the constraints significantly improve when compared to the single sparsity measurements. As we can see in Fig.~\ref{fig:2D_chex}, the use of additional sparsities indeed breaks the $S_8$ degeneracy. From the values quoted in Table~\ref{tab:chexmate_res}, we notice that accounting for the HE and baryon bias slightly alter the area under the $1\sigma$ credibility contours, with the baryon bias case corresponding to the larger value and that with the intrinsic systematics only corresponding to the smallest value. This is consistent with the difference in amplitude of the systematic shifts between the HE and baryon case respectively. Nevertheless, we can see that such difference have no impact on the marginalized $1\sigma$ errors on the cosmological parameters.

Nevertheless, it is important to notice that such bias effects were estimated using results of N-body/hydro simulations which were not specifically devoted to the study of the halo sparsity. Hence, we advocate for a more in-depth study of the influence of baryonic processes on the mass profile of haloes as traced by sparsity measurements, which we leave for a future study.

Finally, we would like to stress that for this type of cosmological parameter inference to be possible, independent cluster mass measurements at multiple overdensity need to be carried out. This implies adopting new methodologies which abandon the two-parameters NFW fitting profile in favour of more general, non-parametric approaches \citep[see e.g.][for a review]{2013SSRv..177..119E}. Recent examples of these procedures to infer the galaxy cluster mass profile with a non-parametric method have been presented in \citep{2018A&A...617A..64B} to derive NFW-independent sparsity estimates of a sample of high-redshift clusters, and in \citet{2019A&A...628A..86B} to test against the standard forward/backward NFW methods. More recently, \citet{2022arXiv220501110E} have developed a forward non-parametric method to derive mass profiles independent from any functional form of the potential. Weak-lensing observations can also provide non-parametric estimates of the mass profile (and consequently of the cluster sparsity) through mass aperture measurements \citep[see e.g.][for a recent study]{2022arXiv220316379D}.

\section{Summary and Discussion}\label{conclusions}
The gravitational mass assembly process that leads to the formation of dark matter haloes, which host galaxy groups and clusters, imprints cosmological information on the halo mass profiles. This can be retrieved through measurements of the halo sparsity, i.e. the ratio between  halo masses enclosing two different overdensities, that has been shown to provide a non-parametric proxy for the halo internal mass distribution \citep{Balmes2014}. In the past few years, cosmological constraints have been inferred from measurements of the average sparsity of galaxy cluster samples using hydrostatic equilibrium masses at $\Delta=500\rho_c$ and $1000\rho_c$ from X-ray observations \citep{Corasaniti2018} and weak lensing masses at $\Delta=200\rho_c$ and $500\rho_c$ \citep{Corasaniti2021}. However, cosmological information is encoded over the entire halo mass profile, rather than at only two overdensities. 

Here, we have investigated the use of multiple sparsity measurements from halo mass estimates at several overdensity as a probe of the cosmological imprint on the halo mass profile. For this purpose, we have analysed N-body halo catalogues from the Raygal and M2Csims simulations and estimated the correlation among different sparsities as a function of redshift. In particular, we have focused on halo masses evaluated at four different overdensities, thus allowing to estimate a total of six sparsities. Interestingly, we find that, among these sparsities, those associated with the mass distribution in distinct spherical halo shells are not highly correlated. Thus, indicating that there is additional cosmological information encoded in the average halo mass profile, which can be exploited through multiple sparsity measurements. In contrast, sparsities obtained using mass estimates derived from the NFW best-fitting density profile to the N-body haloes result in correlations that are close to unity and significantly different from those inferred from the analysis of the spherical overdensity N-body halo masses. This suggests that imposing a NFW profile to haloes performs a strong compression that misses cosmological information imprinted on different regions of the halo mass profile. 

To assess the constraining power of multiple sparsity measurements, we have performed a MCMC likelihood analysis of synthetic generated datasets from the Raygal and M2Csims simulations consisting of different number of sparsities, from a single sparsity case up to a total of six, and inferred cosmological parameter constraints on $\Omega_m$ and $\sigma_8$. We find the constraints to improve as the number of sparsities used increases, with a maximal effect for the case with four sparsities. Instead, the constraints saturates beyond the four sparsity case, which suggest that the additional sparsity estimates only provide redundant information. We have also performed a forecast analysis for a synthetic dataset of four average sparsity measurements generated assuming the characteristic of a realistic cluster sample such as that from the CHEX-MATE project. We have inferred cosmological parameter constraints for different errors assumptions, including the impact of systematic effects on sparsities due to baryons or deviations from the hydrostatic equilibrium, from mass bias estimates obtained from past studies using N-body/hydrodynamical simulations. The results show that these effects only mildly impact the cosmological parameter inference, although dedicated numerical studies are still needed to derive more accurate predictions for baryon systematics on sparsity measurements. 

It has been long considered that cosmological information encoded in the halo density profile can be retrieved through measurements of the concentration-mass relation \citep[see e.g.][]{2010A&A...524A..68E}. The observational challenges posed by the necessity of having accurate measurements of the concentration parameter of galaxy clusters has been the primary limitation for the use of such an approach \citep[][]{2010MNRAS.406..434M,2011MNRAS.416.2539K,2015MNRAS.449.2024S}. 

Our study not only shows that the use of halo sparsity provides a more direct and simpler way to access such information as already discussed in past analyses \citep{Corasaniti2018,Corasaniti2021}, but also that multiple sparsity measurements can fully exploit the cosmological signal imprinted in the mass profile, which would be otherwise missed if the halo density profile was assumed to be NFW. Because of this, we encourage the development of methodologies capable of providing independent mass estimates at different overdensities free of the assumption of the NFW profile. 
  
\section*{Acknowledgements}
The authors are thankful to Romain Teyssier for his role in the development of the M2Csims simulations suite. AMCLB is grateful to Christian Arnold and Baojiu Li for granting access to their Dirac allocation on COSMA for running \textsc{pSOD} on the M2Csims simulations. AMCLB was supported by the French Agence Nationale de la Recherche under grant ANR-11-BS56-015 and by the European Research Council under the European Union Seventh Framework Programme (FP7/2007-2013) / ERC grant agreement number 340519 while conducting the M2Csims simulation programme. AMCLB is currently supported by a fellowship of PSL University at the Paris Observatory. PSC, TR and YR  acknowledge support from the DIM ACAV of the Region Ile-de-France. SE acknowledges financial contribution from the contracts ASI-INAF Athena 2019-27-HH.0, 
``Attivit\`a di Studio per la comunit\`a scientifica di Astrofisica delle Alte Energie e Fisica Astroparticellare''
(Accordo Attuativo ASI-INAF n. 2017-14-H.0), INAF mainstream project 1.05.01.86.10, and from the European Union Horizon 2020 Programme under the AHEAD2020 project (grant agreement n. 871158). This research was supported by the Munich Institute for Astro- and Particle Physics (MIAPP) which is funded by the Deutsche Forschungsgemeinschaft (DFG, German Research Foundation) under Germany's Excellence Strategy - EXC-2094 - 390783311.
The work presented here was granted access to HPC resources of TGCC/CINES through allocations made by GENCI (Grand Equipement National de Calcul Intensif) under the allocations 2016-042287, 2017-A0010402287, 2018-A0030402287, 2019-A0050402287 and 2020-A0070402287 for RayGal and of CINES under allocations 2015-047350, 2016-047350, 2017-A002047350, 2018-A004047350, 2019-A006047350 and 2020-A008047350 for M2Csims, respectively. This work used the DiRAC@Durham facility managed by the Institute for Computational Cosmology on behalf of the STFC DiRAC HPC Facility (www.dirac.ac.uk). The equipment was funded by BEIS capital funding via STFC capital grants ST/P002293/1, ST/R002371/1 and ST/S002502/1, Durham University and STFC operations grant ST/R000832/1. DiRAC is part of the National e-Infrastructure.

\section*{Data Availability}
The data which were used in the study presented here will be made available upon reasonable request to the corresponding authors. Data from the Raygal simulation suite including light-cone halo catalogues accounting for relativistic effects are publicly available at \href{https://cosmo.obspm.fr/public-datasets/}{https://cosmo.obspm.fr/public-datasets/}.


\bibliographystyle{mnras}
\bibliography{bibliography} 


\appendix

\section{Halo Mass Function Best-Fitting Coefficients}\label{app_a}
\begin{figure*}
    \centering
    \includegraphics[width = 0.48\linewidth]{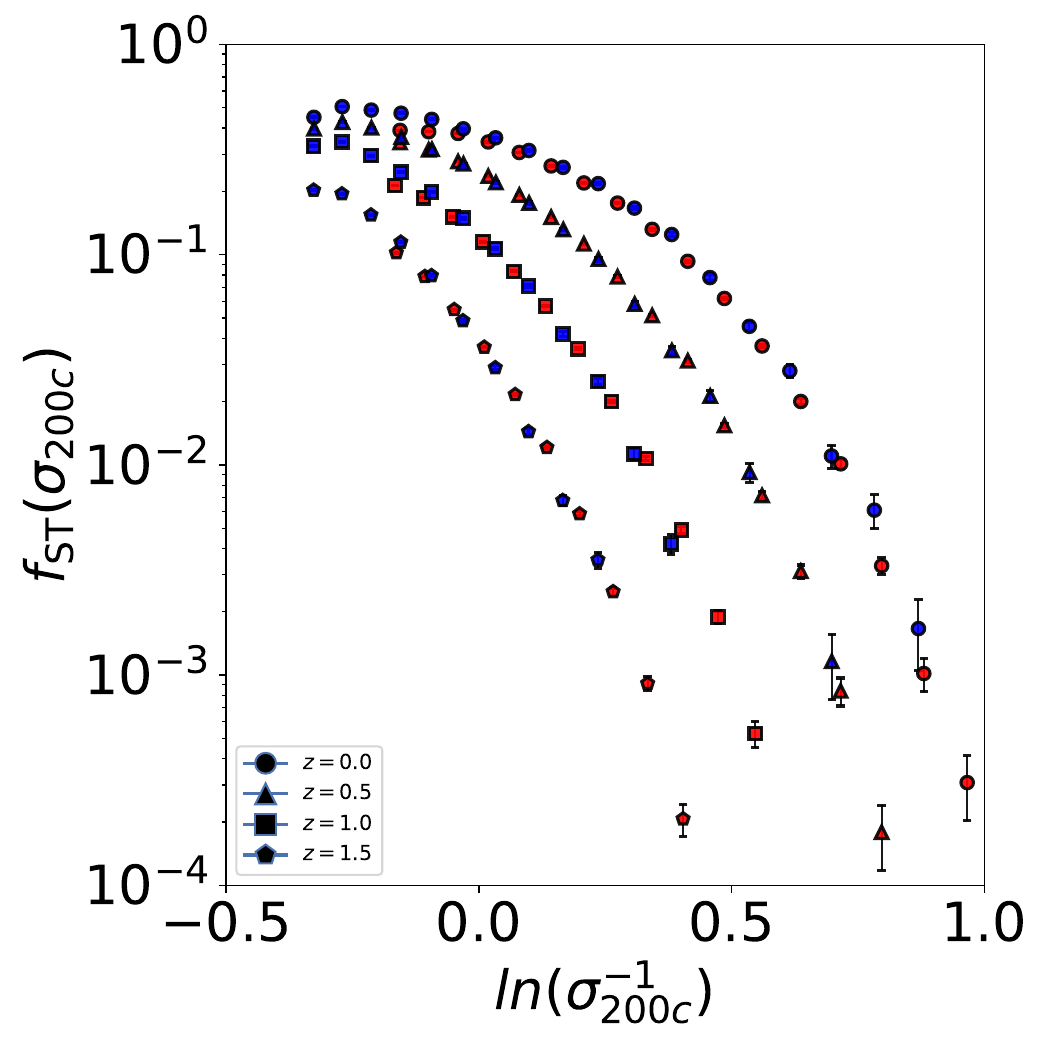}\
    \includegraphics[width = 0.48\linewidth]{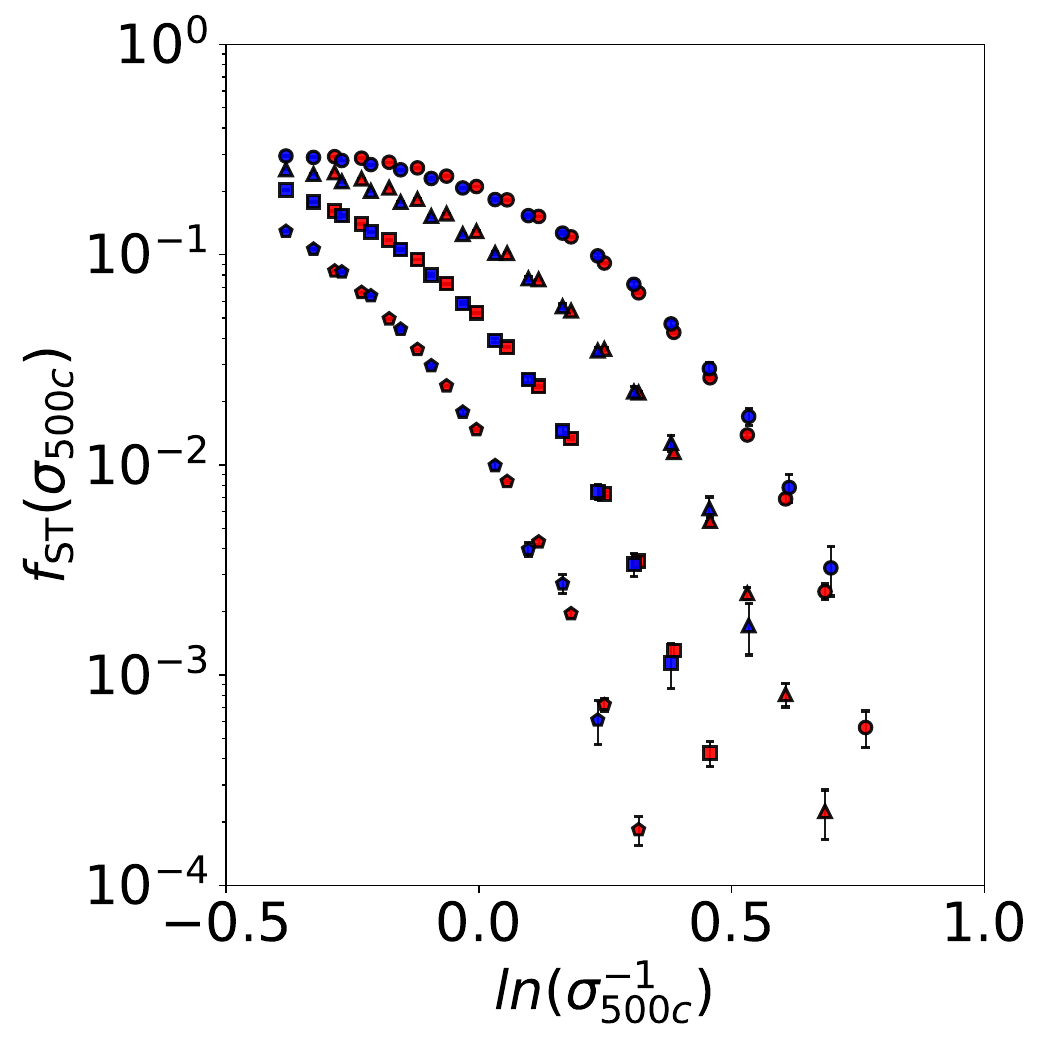}\
    \includegraphics[width = 0.48\linewidth]{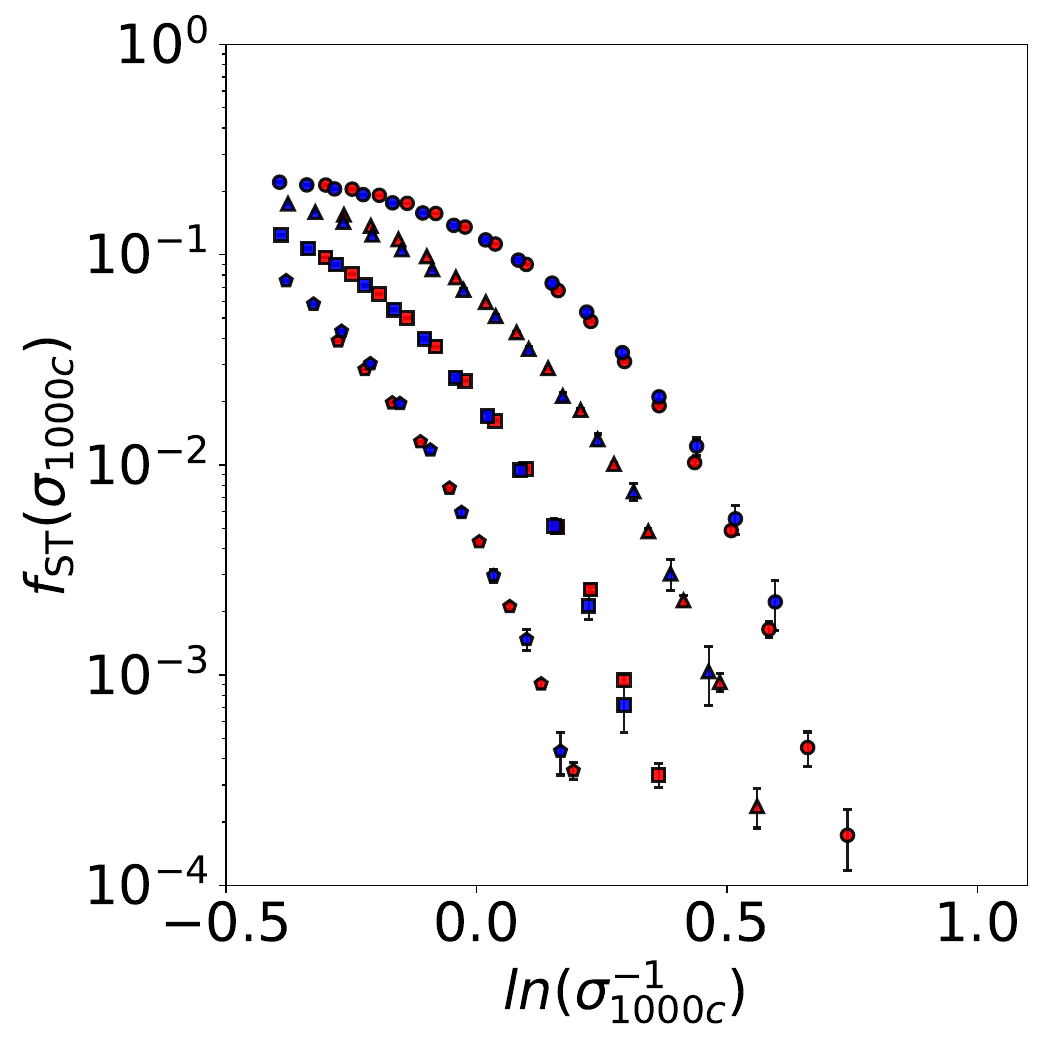}\
    \includegraphics[width = 0.48\linewidth]{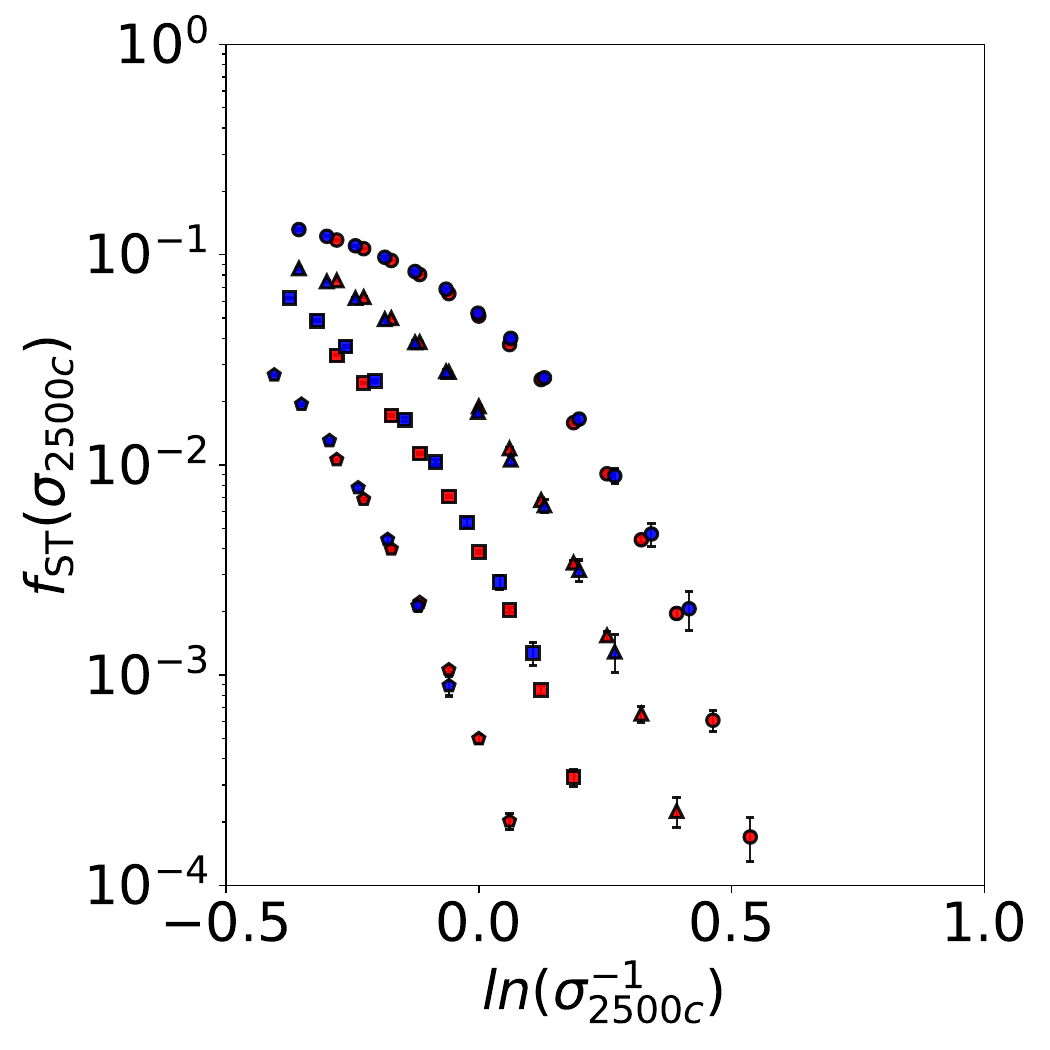}
    \caption{Multiplicity function at $\Delta=200$ (top left panel), $500$ (top right panel), $1000$ (bottom left panel) and $2500$ (bottom right panel) from the Raygal (red points) and M2Csims (blue points) simulations, respectively at $z=0.0$ (circles), $0.5$ (triangles), $1.0$ (squares) and $1.5$ (pentagons). Given the proximity of the simulated cosmologies, the multiplicity functions estimated from the Raygal and M2Csims halo catalogues are consistent with one another within Poisson errors over the common range of masses probed by the simulations.}\label{fig:fst} 
\end{figure*}

We use the halo mass functions from the Raygal and M2Csims halo catalogues to fit the coefficients of the mass function parametrisation given by Eq.~(\ref{st_multiplicity}), which we determine using a Levenberg-Marquardt minimisation scheme. The multiplicity functions at $\Delta=200,500,1000$ and $2500$ estimated from the Raygal and M2Csims halo catalogues are shown in Fig.~\ref{fig:fst}. Given the proximity of the simulated cosmological models, we can see that the estimated multiplicity functions are in good agreement with one another. Notice that since the Raygal simulation probes a slightly larger cosmic volume than the M2Csims simulations, the corresponding multiplicity functions extend over larger $\ln{\sigma^{-1}}$ values. Conversely, the M2Csims simulations have slightly better mass resolution, thus probing smaller $\ln{\sigma^{-1}}$ values than Raygal multiplicity functions.

\begin{figure*}
    \includegraphics[width = 0.5\linewidth]{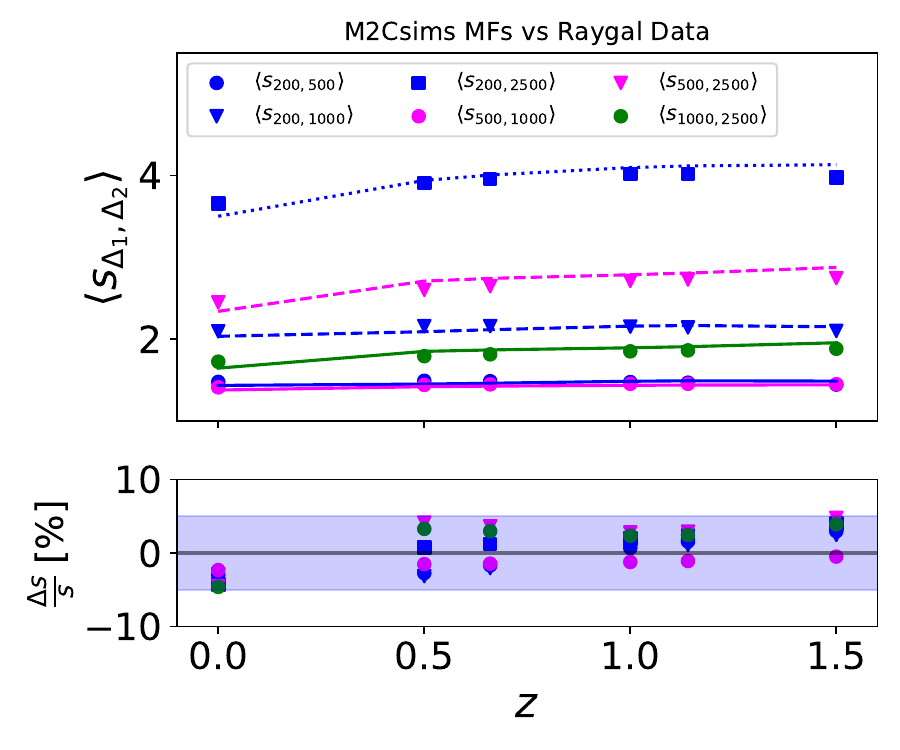}\includegraphics[width = 0.5\linewidth]{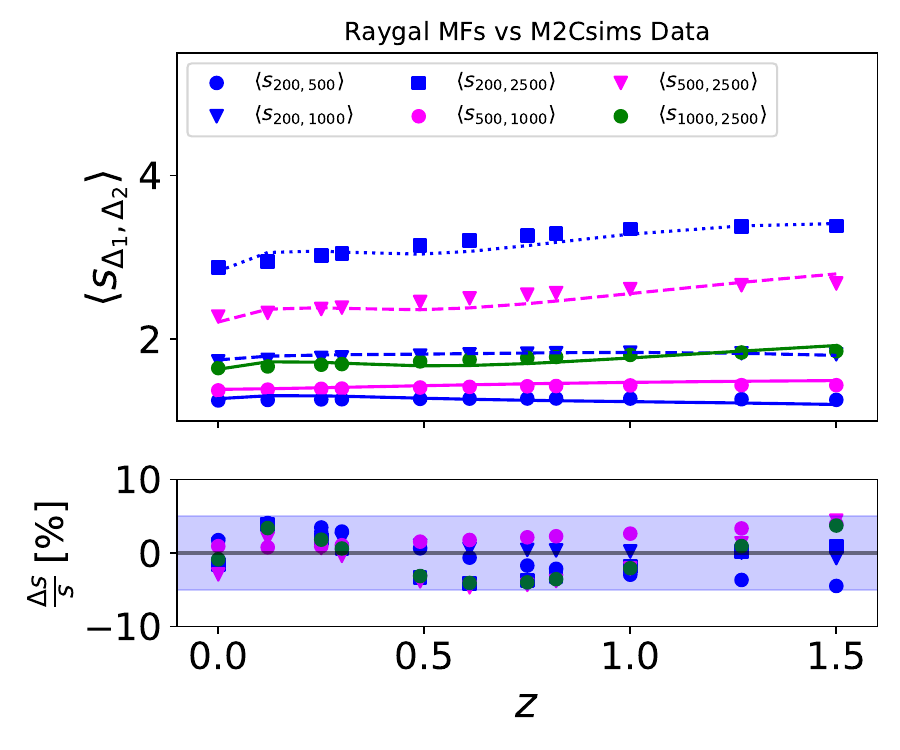}
    \caption{Left panel: average halo sparsity for different overdensity configurations as a function of redshift predicted by solving Eq.~(\ref{spars_mf}) using the M2Csims calibrated multiplicity functions for the Raygal cosmology plotted against the average sparsities from the Raygal halo catalogues. Right panel: average halo sparsities predicted by solving Eq.~(\ref{spars_mf}) using the Raygal calibrated multiplicity functions for the M2Csims cosmology plotted against the average sparsities from the M2Csims halo catalogues. The lower panels show the relative difference between the predictions and the N-body halo catalogue estimates.}\label{fig:av_spars_cosmo_test}
\end{figure*}

In order to predict the average sparsity at redshifts different from those probed by the simulation snapshots using Eq.~(\ref{spars_mf}), we introduce the redshift dependent variable $x=\log_{10}(\Delta/\Delta_{\rm vir}(z))$, where $\Delta_{\rm vir}(z)$ is the virial overdensity as given by the formula derived in \citet{BryanNorman1998}, then following \citet{2016MNRAS.456.2486D}, we parametrise the redshift evolution of the best-fit ST coefficients as a quadratic function of $x$: 
\begin{equation}
\theta_{\Delta}= c_0 + c_1\cdot x + c_2\cdot x^2,
\end{equation}
where $\theta_{\Delta}=\{A_{\Delta},a_{\Delta},p_{\Delta}\}$. In Tab.~\ref{tab:st_coeff_x_raygal} and Tab.~\ref{tab:st_coeff_x_m2csims}, we quote the values of the quadratic parametrisations for the ST-coefficients.

In Fig.~\ref{fig:av_spars_cosmo_test} we plot the predictions from Eq.~(\ref{spars_mf}) for the Raygal cosmology using the M2Csims calibrated multiplicity functions against the average halo sparsity estimates from the Raygal halo catalogues (left panel) and the predictions for the M2Csims cosmology using the Raygal calibrated multiplicity functions against the average halo sparsity estimates from the M2Csims halo catalogues (right panel). As we can see differences are $\lesssim 5\%$, consistent with those shown in Fig.~\ref{fig:av_spars}.

\begin{table}
\centering
\caption{Coefficients of the quadratic function of $x$ parametrising the redshift evolution of the ST parameters for the Raygal halo mass functions}
\begin{tabular}{c|c|c|c}
\hline
             & $c_0$ & $c_1$ & $c_2$ \\
\hline
 $A_{200c}$  & $-0.134134392$ & $5.018016486$ & $-10.7419621$\\
 $a_{200c}$ & $0.989977371$ & $-0.5291517036$ & $4.430318949$ \\
 $p_{200c}$ & $-1.688969793$ & $10.0338655815$ & $22.480766636$ \\
 \hline
 $A_{500c}$ & $-0.52885598$ & $2.4986942$ & $-1.91309251$ \\
 $a_{500c}$ & $-0.408103836$ & $3.5426469558$ & $-1.249831194$ \\
 $p_{500c}$ & $0.380460633$ & $-2.894905482$ & $1.9075409115$ \\
 \hline
  $A_{1000c}$ & $-1.20684329$ & $3.07924433$ & $-1.65792698$ \\
 $a_{1000c}$ & $-1.9777704111$ & $4.8809416194$  & $-1.1738202399$\\
 $p_{1000c}$ & $1.856153925$& $-4.089186069$& $1.5184593426$ \\
 \hline
  $A_{2500c}$ & $7.1510462212$ & $-10.4752748702$ & $3.96694467$ \\
 $a_{2500c}$ & $-9.5732094279$ & $13.1786821425$ & $-3.257783121$\\
 $p_{2500c}$ & $52.2138579936$ & $-77.6915082432$ & $28.3514283168$ \\
 \hline
\end{tabular}\label{tab:st_coeff_x_raygal}
\end{table}

\begin{table}
\centering
\caption{As in Tab.~\ref{tab:st_coeff_x_raygal} but for the M2Csims halo mass functions.}
\begin{tabular}{c|c|c|c}
\hline
             & $c_0$ & $c_1$ & $c_2$ \\
\hline
 $A_{200c}$  & $-0.1392442736$ & $5.2091790188$ & $-11,1511797962$\\
 $a_{200c}$ & $0.9996830315$ & $-0.534339383$ & $4.4737534485$ \\
 $p_{200c}$ & $-1.8134201988$ & $10.7732030454$ & $-24.1372441776$ \\
 \hline
 $A_{500c}$ & $-1.88354029$ & $7.56649037$ & $-6.128838$ \\
 $a_{500c}$ & $1.32586081$ & $-2.18316668$ & $2.90408494$ \\
 $p_{500c}$ & $-6.74606335$ & $21.7532933$ & $-17.97632164$ \\
 \hline
  $A_{1000c}$ & $-5.13368556$ & $11.95696884$ & $-6.5251012$ \\
 $a_{1000c}$ & $3.98834671$ & $-7.94206751$  & $5.54853335$\\
 $p_{1000c}$ & $-20.43015395$& $45.10712761$& $-25.21860361$ \\
 \hline
  $A_{2500c}$ & $-8.03447658$ & $12.44996314$ & $-4.704367$ \\
 $a_{2500c}$ & $3.06066127$ & $-4.70146171$ & $3.01797036$\\
 $p_{2500c}$ & $-42.49723101$ & $64.44575098$ & $-24.79541975$ \\
 \hline
\end{tabular}\label{tab:st_coeff_x_m2csims}
\end{table}

\section{Sparsity Correlation Coefficient Fitting Functions}\label{app_b}
The redshift evolution of the average sparsity correlation coefficients shown in Fig.~\ref{fig:corr} is well approximated by a linear relation:
\begin{equation}
    r_{s_1,s_2}(z) = q + m\cdot z,
\end{equation}
where the coefficients of the linear regression are given in Table~\ref{r_coeff_raygal} and \ref{r_coeff_m2csims} for the Raygal and M2Csims halo catalogues respectively.

\begin{table}
\centering
    \caption{Linear regression parameters of sparsity correlation coefficients from the Raygal halo catalogues.}
    \begin{tabular}{ccc}
    \hline
    Raygal & $m$ & $q$ \\
    \hline
    $r_{s_{200,500},s_{200,1000}}$ &  $-0.032\pm 0.005$ & $0.898\pm 0.005$\\
    $r_{s_{200,500},s_{200,2500}}$ & $-0.14 \pm 0.01$ & $0.71\pm 0.01$ \\
    $r_{s_{200,500},s_{500,1000}}$ & $-0.08\pm 0.01$ & $0.48+\-0.01$ \\
    $r_{s_{200,500},s_{500,2500}}$ & $-0.19\pm0.01$ & $0.37\pm 0.01$\\
    $r_{s_{200,500},s_{1000,2500}}$ & $-0.22\pm 0.01$ & $0.24\pm 0.01$\\
    \hline
    $r_{s_{200,1000},s_{200,2500}}$ & $-0.081\pm 0.002$ & $0.902\pm 0.002$ \\
    $r_{s_{200,1000},s_{500,1000}}$ & $-0.011\pm 0.002$ & $0.810\pm 0.002$ \\
    $r_{s_{200,1000},{s_{500,2500}}}$ & $-0.132\pm 0.003$ &
    $0.666\pm 0.003$ \\
    $r_{s_{200,1000},s_{1000,2500}}$ & $-0.188\pm 0.004$ &
    $0.461\pm 0.004$ \\
    \hline
    $r_{s_{200,2500},s_{500,1000}}$ & $-0.030\pm 0.002$ &
    $0.855\pm 0.002$ \\
    $r_{s_{200,2500},s_{500,2500}}$ & $-0.003\pm 0.003$ &
    $0.905\pm 0.003$ \\
    $r_{s_{200,2500},s_{1000,2500}}$ & $-0.024\pm 0.005$ &
    $0.784\pm 0.004$ \\
    \hline
    $r_{s_{500,1000},s_{500,2500}}$ & $-0.076\pm 0.002$ &
    $0.858\pm 0.001$ \\
    $r_{s_{500,1000},s_{1000,2500}}$ & $-0.127\pm 0.003$ &
    $0.619\pm 0.003$ \\
    \hline
    $r_{s_{500,2500},s_{1000,2500}}$ & $-0.006\pm 0.001$ &
    $0.927\pm 0.001$ \\
    \hline
    \end{tabular}\label{r_coeff_raygal}
\end{table}

\begin{table}
\centering
    \caption{Linear regression parameters of sparsity correlation coefficients from the M2Csims halo catalogues.}
    \begin{tabular}{ccc}
    \hline
    M2Csims & $m$ & $q$ \\
    \hline
    $r_{s_{200,500},s_{200,1000}}$ &  $-0.028\pm 0.003$ & $0.901\pm 0.002$\\
    $r_{s_{200,500},s_{200,2500}}$ & $-0.134\pm 0.006$ & $ 0.726\pm 0.005$ \\
    $r_{s_{200,500},s_{500,1000}}$ & $-0.079\pm 0.004$ & $0.48+\-0.01$ \\
    $r_{s_{200,500},s_{500,2500}}$ & $-0.182\pm 0.003$ & $0.395\pm 0.003$\\
    $r_{s_{200,500},s_{1000,2500}}$ & $-0.212\pm 0.004$ & $0.262\pm 0.003$\\
    \hline
    $r_{s_{200,1000},s_{200,2500}}$ & $-0.074\pm 0.001$ & $0.909\pm 0.001$ \\
    $r_{s_{200,1000},s_{500,1000}}$ & $-0.014\pm 0.003$ & $0.826\pm 0.003$ \\
    $r_{s_{200,1000},{s_{500,2500}}}$ & $-0.123\pm 0.003$ &
    $0.685\pm 0.003$ \\
    $r_{s_{200,1000},s_{1000,2500}}$ & $-0.170\pm 0.004$ &
    $0.478\pm 0.003$ \\
    \hline
    $r_{s_{200,2500},s_{500,1000}}$ & $-0.028\pm 0.002$ &
    $0.870\pm 0.002$ \\
    $r_{s_{200,2500},s_{500,2500}}$ & $-0.003\pm 0.003$ &
    $0.910\pm 0.002$ \\
    $r_{s_{200,2500},s_{1000,2500}}$ & $-0.019\pm 0.004$ &
    $0.787\pm 0.003$ \\
    \hline
    $r_{s_{500,1000},s_{500,2500}}$ & $-0.070\pm 0.001$ &
    $0.869\pm 0.001$ \\
    $r_{s_{500,1000},s_{1000,2500}}$ & $-0.112\pm 0.003$ &
    $0.635\pm 0.002$ \\
    \hline
    $r_{s_{500,2500},s_{1000,2500}}$ & $-0.002\pm 0.001$ &
    $0.926\pm 0.001$ \\
    \hline
    \end{tabular}\label{r_coeff_m2csims}
\end{table}


\label{lastpage}
\end{document}